\documentclass[twocolumn]{aastex62} 

\usepackage[caption=false]{subfig}
\usepackage{natbib}
\usepackage{comment} 
\usepackage{csvsimple, array, longtable, booktabs, amsmath, amssymb, bm, hyphenat} 
\usepackage{hyperref}
\hypersetup{ %
                colorlinks=true, %
                pdfauthor={Maurice L. Wilson}, %
                pdftitle={MINERVA's First Radial Velocities}, %
                citecolor=blue, %
                linkcolor=blue %
                }

\usepackage{etoolbox}

\newcommand{\Lfctor}{0.5}

\newcommand{\affilHarvardCfA}{Center for Astrophysics, Harvard \& Smithsonian, 60 Garden St, Cambridge, MA 02138, USA}
\newcommand{\affilUUtah}{Department of Physics and Astronomy, University of Utah, 115 South 1400 East, Salt Lake City, UT 84112, USA}
\newcommand{\affilUPennDept}{The University of Pennsylvania, Department of Physics and Astronomy, Philadelphia, PA 19104, USA}
\newcommand{\affilUFl}{Department of Astronomy, University of Florida, P.O. Box 112055, Gainesville, FL 32611, USA}
\newcommand{\affilMissouri}{Missouri State University, USA}
\newcommand{\affilUSQ}{Centre for Astrophysics, University of Southern Queensland, USQ Toowoomba, QLD 4350, Australia}
\newcommand{\affilCallaghan}{KiwiStar Optics, Callaghan Innovation, New Zealand}
\newcommand{\affilThacher}{The Thacher School, 5025 Thacher Road, Ojai, CA 93023, USA}
\newcommand{\affilBU}{Department of Astronomy \& The Institute for Astrophysical Research, Boston University, 725 Commonwealth Ave., Boston, MA 02215, USA}
\newcommand{\affilMontana}{Department of Physics and Astronomy, University of Montana, 32 Campus Drive, No. 1080, Missoula, MT 59812, USA}

\accepted{2019 July 19}
\submitjournal{PASP}

\begin{document}

\title{First radial velocity results from the MINiature Exoplanet Radial Velocity Array (MINERVA) }
\shorttitle{MINERVA's First Radial Velocities}
\shortauthors{Wilson et al.}

\author[0000-0003-1928-0578]{Maurice L. Wilson} 
\affiliation{\affilHarvardCfA}

\author[0000-0003-3773-5142]{Jason D.\ Eastman} 
\affiliation{\affilHarvardCfA} 

\author[0000-0003-1012-4771]{Matthew A. Cornachione} 
\affiliation{\affilUUtah}
\affiliation{Department of Physics, United States Naval Academy, 572C Holloway Rd., Annapolis, MD 21402, USA}

\author[0000-0002-6937-9034]{Sharon X. Wang} 
\affiliation{Department of Terrestrial Magnetism, Carnegie Institution for Science, 5241 Broad Branch Road, NW, Washington, DC 20015, USA}

\author[0000-0001-9397-4768]{Samson A. Johnson} 
\affiliation{Department of Astronomy, The Ohio State University, 140 West 18th Avenue, Columbus, OH 43210, USA}

\author{David H. Sliski} 
\affiliation{\affilUPennDept}

\author[0000-0001-5255-8274]{William J. Schap III} 
\affiliation{\affilUFl} 

\author[0000-0002-8537-5711]{Timothy D. Morton} 
\affiliation{\affilUFl}
\affiliation{Center for Computational Astrophysics, Flatiron Institute, 162 5th Ave, New York, NY 10010, USA}

\author[0000-0002-1704-6289]{John Asher Johnson} 
\affiliation{\affilHarvardCfA}

\author[0000-0002-8041-1832]{Nate McCrady} 
\affiliation{\affilMontana}

\author[0000-0001-6160-5888]{Jason T. Wright} 
\affiliation{Department of Astronomy and Astrophysics and Center for Exoplanets and Habitable Worlds, The Pennsylvania State University, University Park, PA 16802, USA}

\author[0000-0001-9957-9304]{Robert A. Wittenmyer} 
\affiliation{\affilUSQ}

\author[0000-0002-8864-1667]{Peter Plavchan} 
\affiliation{Department of Physics and Astronomy, George Mason University, Fairfax, VA 22030, USA}

\author[0000-0002-6096-1749]{Cullen H. Blake} 
\affiliation{\affilUPennDept}

\author[0000-0002-9486-818X]{Jonathan J. Swift} 
\affiliation{\affilThacher}

\author[0000-0003-1341-5531]{Michael Bottom} 
\affiliation{Jet Propulsion Laboratory, California Institute of Technology, Pasadena, CA 91109, USA}


\author[0000-0002-6525-7013]{Ashley D. Baker} 
\affiliation{\affilUPennDept}

\author[0000-0001-6027-5370]{Stuart I. Barnes} 
\affiliation{Stuart Barnes Optical Design, The Netherlands} 

\author{Perry Berlind} 
\affiliation{\affilHarvardCfA} 

\author{Eric Blackhurst} 
\affiliation{PlaneWave Instruments Inc., 1819 Kona Drive, Rancho Dominguez, CA 90220, USA}

\author[0000-0002-9539-4203]{Thomas G. Beatty} 
\affiliation{Department of Astronomy and Steward Observatory, University of Arizona, Tucson, AZ 85721, USA}

\author[0000-0002-9836-603X]{Adam S. Bolton} 
\affiliation{\affilUUtah}
\affiliation{National Optical Astronomy Observatory, 950 North Cherry Ave, Tucson, AZ 85719, USA}

\author{Bryson Cale} 
\affiliation{\affilMissouri} 

\author{Michael L. Calkins} 
\affiliation{\affilHarvardCfA}

\author{Ana Col\'{o}n} 
\affiliation{6127 Wilder Lab, Department of Physics \& Astronomy, Dartmouth College, Hanover, NH 03755, USA}

\author{Jon de Vera} 
\affiliation{Las Cumbres Observatory Global Telescope Network, 6740 Cortona Dr. Suite 102, Goleta, CA 93117, USA}

\author[0000-0002-9789-5474]{Gilbert Esquerdo}  
\affiliation{\affilHarvardCfA}

\author[0000-0002-7061-6519]{Emilio E. Falco} 
\affiliation{\affilHarvardCfA} 

\author{Pascal Fortin}  
\affiliation{\affilHarvardCfA} 

\author[0000-0003-1361-985X]{Juliana Garcia-Mejia} 
\affiliation{\affilHarvardCfA}

\author{Claire Geneser} 
\affiliation{\affilMissouri} 

\author{Steven R. Gibson} 
\affiliation{Space Sciences Laboratory, University of California, Berkeley, CA 94720, USA}

\author{Gabriel Grell} 
\affiliation{Department of Astronomy, University of Maryland, College Park, MD 20771, USA}

\author{Ted Groner} 
\affiliation{\affilHarvardCfA} 

\author{Samuel Halverson} 
\affiliation{MIT Kavli Institute for Astrophysics and Space Research, 77 Massachusetts Avenue, 37-241, Cambridge, MA 02139, USA}
\affiliation{NASA Sagan Postdoctoral Fellow}

\author{John Hamlin}  
\affiliation{\affilCallaghan}

\author{M. Henderson} 
\affiliation{Ann and H.J. Smead Department of Aerospace Engineering Sciences, University of Colorado Boulder, Boulder, CO, 80309-0429, USA}

\author[0000-0002-1160-7970]{J. Horner} 
\affiliation{\affilUSQ}

\author{Audrey Houghton}  
\affiliation{\affilMontana} 

\author{Stefaan Janssens} 
\affiliation{\affilCallaghan}

\author{Graeme Jonas} 
\affiliation{\affilCallaghan}

\author{Damien Jones} 
\affiliation{Prime Optics, Queensland, Australia}

\author{Annie Kirby} 
\affiliation{Las Cumbres Observatory Global Telescope Network, 6740 Cortona Dr. Suite 102, Goleta, CA 93117 USA}

\author{George Lawrence}  
\affiliation{\affilThacher}

\author{Julien Andrew Luebbers} 
\affiliation{\affilThacher}

\author[0000-0002-0638-8822]{Philip S. Muirhead} 
\affiliation{\affilBU}

\author[0000-0001-6145-5859]{Justin Myles} 
\affiliation{Department of Physics, Stanford University, 382 Via Pueblo Mall, Stanford, CA 94305, USA}
\affiliation{Kavli Institute for Particle Astrophysics \& Cosmology, P. O. Box 2450, Stanford University, Stanford, CA 94305, USA}

\author[0000-0001-8838-3883]{Chantanelle Nava} 
\affiliation{\affilHarvardCfA}

\author{Kevin O Rivera-Garc\'{i}a} 
\affiliation{University of Puerto Rico, R\'{i}o Piedras Campus}

\author{Tony Reed} 
\affiliation{\affilCallaghan}

\author{Howard M. Relles} 
\affiliation{\affilHarvardCfA} 

\author[0000-0002-0387-370X]{Reed Riddle} 
\affiliation{California Institute of Technology, 1200 E. California Blvd., Pasadena, CA  91125, USA}

\author[0000-0003-1639-510X]{Connor Robinson} 
\affiliation{\affilBU}

\author{Forest Chaput de Saintonge}  
\affiliation{\affilMontana}

\author{Anthony Sergi}  
\affiliation{\affilMontana}

\correspondingauthor{Maurice L. Wilson}
\email{maurice.wilson@cfa.harvard.edu}

\begin{abstract}
The MINiature Exoplanet Radial Velocity Array (MINERVA) is a dedicated observatory of four 0.7 m robotic telescopes fiber-fed to a KiwiSpec spectrograph.  The MINERVA mission is to discover super-Earths in the habitable zones of nearby stars. This can be accomplished with MINERVA's unique combination of high precision and high cadence over long time periods. In this work, we detail changes to the MINERVA facility that have occurred since our previous paper. We then describe MINERVA's robotic control software, the process by which we perform 1D spectral extraction, and our forward modeling Doppler pipeline. In the process of improving our forward modeling procedure, we found that our spectrograph's intrinsic instrumental profile is stable for at least nine months. Because of that, we characterized our instrumental profile with a time-independent, cubic spline function based on the profile in the cross dispersion direction, with which we achieved a radial velocity precision similar to using a conventional ``sum-of-Gaussians'' instrumental profile: 1.8~m s$^{-1}$ over 1.5 months on the RV standard star HD~122064. Therefore, we conclude that the instrumental profile need not be perfectly accurate as long as it is stable. In addition, we observed 51 Peg and our results are consistent with the literature, confirming our spectrograph and Doppler pipeline are producing accurate and precise radial velocities. 
 
 
 
\end{abstract}

\section{Introduction}

The discovery of the first planets orbiting solar-type stars was achieved using Doppler spectroscopy \citep{Campbell.1988,Latham.1989,Mayor.1995}. As the first exoplanet detections and confirmations were made, Doppler spectroscopy instruments gradually improved from attaining a radial velocity (RV) precision of $\sim$15~m~s$^{-1}$~\citep{Campbell.1988} to $\sim$3~m~s$^{-1}$~\citep{Butler.1996} thanks to the advent of the iodine absorption cell technique.  Two decades later, the next generation of precision RV instruments aims for instrumental stability at the 30~cm~s$^{-1}$ level \citep{Wright.2017}. However, our sensitivity to exoplanets is likely limited by stellar activity at the $\sim$1~m~s$^{-1}$ level for most stars~\citep[e.g.,][]{Saar.1997,Haywood.2016}.  Detections below this level will not be achieved until astrophysical noise sources are understood as well as sources of instrumental noise.  Observing with high cadence throughout a planet's full orbit may allow us to understand and correct for non-planetary RV signals induced by stellar activity~\citep{OToole.2008, Pepe.2011, Dumusque.2012}.

The MINiature Exoplanet Radial Velocity Array (MINERVA) is a dedicated observatory aiming for both high cadence and high precision RV measurements \citep{Swift.2015}. It is a robotic array of four 0.7~m telescopes located on Mt. Hopkins in Arizona.  The MINERVA mission ultimately has two objectives.  

The primary science objective is to detect and characterize super-Earths in the habitable zones of nearby stars.  Our RV target list is a subset of the targets monitored during the NASA/UC $\eta_\oplus$ Survey performed by the California Planet Search (CPS) group at the Keck Observatory using the HIRES spectrograph~\citep{Howard.2009}.  Out of the 230 GKM stars they surveyed, 166 are considered chromospherically quiet~\citep{Wright.2004,Isaacson.2010}.  The MINERVA RV target list consists of 125 of the brightest ($V \lesssim 8$) chromospherically quiet stars from their survey that can be observed from southern Arizona.  With MINERVA's effective aperture of 1.4~m and use of the NASA/UC $\eta_\oplus$ targets,  the RV precision goal of the MINERVA mission was set to detect planets at the 80~cm~s$^{-1}$ level~\citep{Swift.2015}.  At this level, we plan to characterize super-Earths while providing insight into the importance of cadence as a tool for understanding the problem of stellar activity. We show that we are about a factor of 2 of that goal in \S \ref{sect: rv standard}, which is already within the top tier of the current generation of precision RV instruments. Coupled with our unmatched observational cadence, we are already operating in a unique parameter space that will enable us to detect new planets and provide valuable insight about the importance of cadence in understanding stellar jitter. We can do this with our cost-efficient, four-telescope, robotic array observing at an unprecedented cadence.  The high cadence is attributed to the autonomous, flexible target scheduling, and quick slewing of the CDK-700 telescopes.  Most importantly, the majority of the robotic array's time is not split between multiple teams or science goals.

The secondary science objective is to search for transits of the super-Earths we find.  This requires a broadband photometric precision of $<$1~mmag in the optical: a goal that has already been demonstrated by \citet{Swift.2015}.  Multiband light curves provide information that otherwise cannot be deduced from Doppler spectroscopy alone.  For example, the minimum mass of the planet can be found from radial velocities, but if the planet happens to transit, the transit photometry can determine the radius and inclination of the planet \citep[see, e.g.][]{Winn.2010}. Therefore, both exoplanet detection methods used together can indicate the true planetary mass and bulk density. 

MINERVA's secondary objective has already contributed to a variety of exoplanet science endeavours~\citep{Swift.2015,Vanderburg.2015,Croll.2017,Lund.2017,Pepper.2017,Rodriguez.2017,Siverd.2018,Labadie.2019}.  Thus, in this work we focus on the commencement of MINERVA's primary objective.  We report our survey performance in \S\ref{sect: status}, the changes to our hardware since our last paper 
in \S\ref{sect: hardware}, the environmental stability of the spectrograph in \S \ref{sec:stability}, our revised telescope control software in \S\ref{sect: software}, our one-dimensional extraction in \S\ref{sec:1d}, our Doppler pipeline in \S\ref{sect: doppler}, our first RV results in \S\ref{sect: RVs}, and our final remarks in \S\ref{sect: conclusion}.

\section{Survey Performance}
\label{sect: status}

Observing at Mt Hopkins is divided naturally into seasons by the July/August monsoon shutdown. The first full-season MINERVA observing campaign in radial velocity survey mode began 2017 September 14 and ran through 2018 June 29. Spectra were obtained on 196 of 293 nights. Weather prevented observations on 44 nights, and 53 nights were spent on engineering or lost to system malfunctions. We obtained 1936 exposures with 4 spectra each of 28 survey target stars, with a maximum of 222 exposures of a single (high decl.) target. Fourteen targets had at least 60 exposures. In addition, we obtained 199 exposures of hot stars, at least one per night, used for spectral calibration. A typical night full of observing led to 12 to 19 exposures (most in December, less on shorter nights) of 8 to 10 target stars. The open shutter fraction was highly variable at the beginning of the season, but stabilized at $\sim69$\% after implementation of the autonomous scheduler in late October. Given the rapid slewing and settling time of our telescopes, the majority of the overhead per spectrum was the result of robotic target acquisition on the fiber tip.

The 2018-19 observing campaign began 2018 October 15 and is in progress at the time of writing this manuscript. Through 2019 March 31, spectra have been obtained on 107 of 168 nights, with 35 nights lost to weather and 26 spent on engineering or lost to system malfunctions. We have obtained 1455 exposures with 4 spectra each of 19 survey target stars, a 32\% increase in spectra obtained over the same period from the previous season. Twelve targets have at least 60 exposures thus far. In addition we have obtained 137 exposures of hot calibration stars. Changes in our acquisition algorithm have reduced the overhead, resulting in an average open shutter fraction of 86\% since 2018 November. Historically, April through June provide very reliable weather at the site (in 2018-2019 we lost only 3 of these 91 nights to weather), and we anticipate that this observing season will lead to a larger set of RV data than the previous season --- 43\% of our 2017-18 spectra were obtained in April--June.

\section{Hardware}\label{sect: hardware}

The overall hardware design for the MINERVA facility has remained largely unchanged from that described in \citet{Swift.2015}.  However, we have made several changes to improve our science capability, which we discuss in detail below.

\subsection{Fiber Acquisition Unit (FAU) Cameras}
\label{sect:zwo}
The SBIG ST-i cameras originally used in the FAUs had two major problems. First, their small field of view (3$^{'}$.6 x 2$^{'}$.7) coupled with a surprisingly quick degradation of the telescope pointing meant that we could not blindly point to a target and be confident it would fall on the detector. We had to redo the pointing model weekly to ensure the pointing was sufficient for robust acquisition---a time-consuming task that must be done manually at night, when the telescopes would otherwise be carrying out science observations. The source of the pointing degradation is not clear, and the telescope manufacturers have not seen such degradation for other users, suggesting a problem with the robotic control software that we have not been able to fully investigate.

Second, the SBIGs had a high failure rate. During the initial month-long spectrograph commissioning, three out the four cameras in use experienced critical failures. 

The manufacturers were aware that this problem affected a small batch of cameras and repaired them. After those repairs, the cameras performed better, but over the three years that followed, several more failures occurred.  Given that replacing a failed camera requires a site visit, at a cost significantly greater than that of the cameras themselves, we decided to replace the SBIG cameras with ZWO ASI 174 cameras. 

In the six months of daily use on four telescopes since their installation, we have not had a single camera failure. These cameras have a CMOS detector with 1936 x 1216, 5.86 $\mu m$ pixels, and is similarly priced. This provides us an 8.5' x 5.4' field of view that allows us to robustly acquire our targets despite the pointing degradation of the telescopes. One downside to these cameras is that they are incompatible with our Black Box USB extenders, and so we had to move our control computers into the domes in order to use them.

\subsection{Fiber}
The science fiber was originally purchased from CeramOptec and has a 50 $\mu$m octagonal core and 330 $\mu$m cladding. As is typical for CeramOptec, they actually have two slightly different claddings for octagonal fibers -- one deposited onto the octagonal core to make it circular, and another with the core drilled out that they plug in with the circularized octagonal core. As was the case with our fiber, the boundary between the two claddings can be problematic if the indices are not well matched, since it can guide starlight through the cladding. Light transmitted through the cladding reduces the resolution of the spectrograph and the instrumental profile can vary dramatically as a function of how much light couples with the cladding---both of which are catastrophic for precision RV measurements. In addition, the cladding was too large to pack together at the focal plane with the required core-to-core spacing, and the standard Ferrule Connector (FC) connectors at each end, done by CeramOptec, suffered from severe Focal Ratio Degradation (FRD) and thus led to a major loss in throughput.

As a short-term solution, we re-terminated the original fibers to improve the FRD, and coupled the fibers into a short section of fiber with a 50~$\mu$m circular core and a 125~$\mu$m cladding to remove light from the cladding of the octagonal fiber. One side is butt-coupled to each of the four telescope fibers, and the other four ends combine into a $V$-groove at the spectrograph end, spaced 220~$\mu$m center to center.

As a long-term solution, we ordered a new fiber with a custom preform (a macro-sized piece of glass from which the fiber is drawn), which had a 50~$\mu$m octagonal core and 110~$\mu$m cladding from Polymicro, with the intention of packing seven of them cladding to cladding to allow for future expansion should we decide the significant ($\sim10\%$) crosstalk from such tightly packed fibers is manageable. Polymicro deposits the entire cladding onto the octagonal core to avoid the secondary cladding issue. However, our cladding was much thicker than typical, and during the lengthy deposition, the core melted and mixed into the cladding, creating the same effect as a secondary cladding. Light was still transmitted through the cladding of the fiber. A second attempt was no better, at which point they would not attempt the expensive process again. We could not afford a thinner cladding because we split the expensive custom preform with two other groups that required a thicker cladding.

So, our short term solution has become our final fiber solution. While the butt-couple is lossy, it makes the installation easier, it provides an easy point to add a double scrambler if we decide it is necessary in the future, and a change in the fiber geometry improves the near-field scrambling \citep{Halverson.2015}. Meanwhile, MINERVA Australis \citep{Addison.2019} has created a fiber similar to our desired long-term solution, with enough for a spare if we decide it is worth the effort to replace in the future.

\subsection{Spectrograph}
The KiwiSpec spectrograph was installed in 2015 December, and is a commercial adaptation of the spectrograph designed and described in \citet{Barnes.2012} and \citet{Gibson.2012}, with a new camera designed by Prime Optics\footnote{http://www.primeoptics.com.au/}. With a few exceptions highlighted below, it is as we described in \citet{Swift.2015}. Instead of the simultaneous etalon or thorium argon wavelength reference described in \citet{Swift.2015}, we used the thorium argon lamp during the installation as a rough wavelength solution and now rely solely on the iodine to provide the exact wavelength solution. The simultaneous wavelength reference in addition to iodine is unnecessary, and removing them allowed us to reduce the scattered light and increase the spacing between fibers, reducing the crosstalk to $\lesssim 0.1\%$. 

We determined our total system throughput to be $\sim5\%$ using Doppler Tomography observations of KELT-24b without the iodine cell \citep{Rodriguez.2019}. We computed the expected flux from the $V$=8.33 host star between 6175-6185 \AA \ and compared it to the actual flux in the extracted 1D spectrum at the same bandpass (at the peak of the blaze). This throughput estimate includes all losses, including the atmosphere, coatings, beam splitter, fiber coupling, Echelle, and charge-coupled device (CCD).

Light from each of the four telescopes is focused directly onto our 50~$\mu$m octagonal fibers at $f$/6.6 in our FAU (see \S \ref{sect:zwo}). Three meters before the spectrograph entrance, each of the fibers are butt-coupled to circular 50~$\mu$m fibers, which are then arranged into a V-groove at the entrance to the spectrograph, separated by 220~$\mu$m center to center. The light exits the four fibers and is collimated. A pupil mask truncates the beam to ensure the beam is precisely $f$/6, allowing for some focal ratio degradation within the fibers. The collimated light travels to the iodine stage, where the iodine gas cell can be moved into or out of the beam. The fibers are then re-imaged onto the entrance slit, and the light follows the path to the detector shown in Figure 1 of \citet{Barnes.2012}.

\begin{figure}
    \centering
    \includegraphics[width=1\linewidth, trim=0cm 0cm 0cm 0.8cm, clip=true]{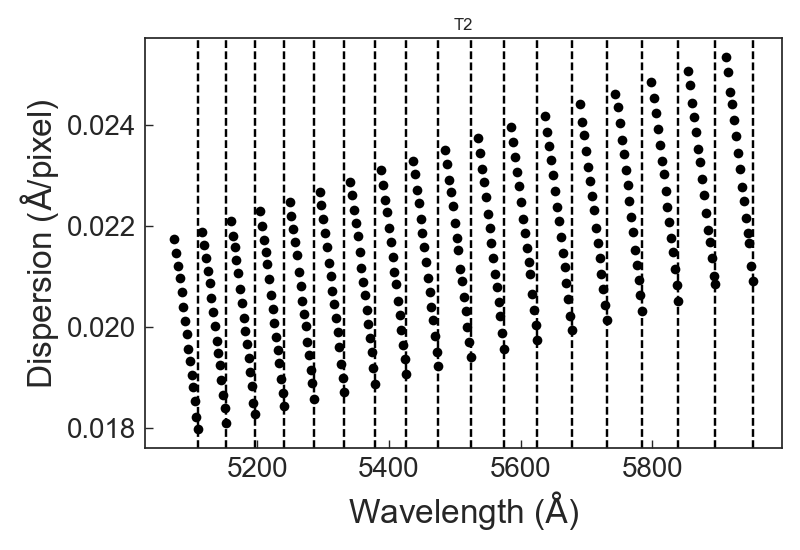}
    \caption{The dispersion fitted for each chunk (black points) that we use to extract the radial velocity as a function of wavelength, for a representative night on a representative telescope. The variance of the dispersion over time and for different telescopes is much much smaller than the size of the data points in the figure. The vertical dashed lines represent the edge of each order.}
    \label{fig:dispersion}
\end{figure}

We empirically determined the resolving power per resolution element of the spectrograph by forward modeling high signal to noise spectra taken of the daytime sky and numerically solving for the FWHM of the fitted IP for each chunk. As expected, there is a slight wavelength dependence in our resolving power. The best-fit line of the resolving power per resolution element as a function of wavelength is $R=84,000 + (\lambda - 5500$\AA$)\times10/$\AA, in good agreement with our theoretical expectation. We also plot the best-fit dispersion per resolution element as a function of wavelength for each chunk in Figure~\ref{fig:dispersion} for a representative night on a representative telescope.

\subsection{Exposure Meter}
We have always had an exposure meter inside the spectrograph that picks off the reflection of the nearly collimated beam from the vacuum window. We have since added a $V$-band filter to approximate the bandpass of the spectrograph. The major downside to this design is that the exposure meter reports an average flux from all four telescopes, so we cannot use it to apply a per-telescope barycentric correction. Instead, we use the guide images from the FAU (at $\sim$5~s cadence), with an aperture the size of the fiber drawn around the measured fiber position to determine the relative flux during the exposure for each telescope. We have confirmed that, when only one telescope is used, we can use the FAU guide images reproduce the relative exposure meter flux to within the uncertainties, in order to compute a per-telescope barycentric correction.

\subsection{Backlight}

The FAU design described in \S \ref{sect:zwo} and \citet{Swift.2015} flexes depending on its rotation angle and the telescope's altitude. This flexing causes the apparent position of the fiber on the acquisition camera to move by $\sim10$~$\mu$m over the sky---or 20\% of the fiber diameter, which would be a significant source of light-loss if left uncorrected. We knew this would be a problem and the FAU was designed to be able to locate the fiber tip on the acquisition unit by backlighting the fiber, but we had not yet fully fleshed out a solution at the time we wrote the \citet{Swift.2015} paper. We considered using the exposure meter to refine the star's position, but that would dramatically increase our acquisition time since it would have to be done serially with each telescope and it can be difficult to make such a procedure robust during variable weather conditions. Ultimately, we added a disk of LEDs that swings in front of the $V$-groove to illuminate the fibers from inside the spectrograph. We do this before each exposure to refine the reference pixel to move the star to, and after to evaluate how much throughput might have been lost due to drift during the exposure. By evaluating a large number of these backlight images, we may be able to map the flexure and eliminate this step and/or compensate for drift during an exposure in the future.

\subsection{Slit flat}
 
Because the fibers do not provide much signal to noise in the wings of their profile, it is difficult to determine the pixel-to-pixel variations in the spectrograph detector with flat fields illuminated through the fiber. We added a light, mounted on the iodine stage, that illuminates a slit where the fibers are re-imaged. We use this flat field to correct for the pixel to pixel variations in the detector, as described in \S \ref{sec:1d}. The flat field lamp simply shines onto the entrance slit, with no attempt to match the $f$/6 science beam. We see no significant scattered light contamination with this approach, but we have yet to perform a detailed investigation.

While our iodine cell was designed to have counter-rotated wedges to eliminate fringing with minimal beam deflection, we believe the parallelism of the iodine cell faces was not within specification. As a result, when the iodine cell is in place, the position of the fibers shifts by almost the entire diameter of the fiber in the dispersion direction. Originally, the slit was only slightly oversized relative to the fiber size. We replaced it with a much wider slit to accommodate both the undeflected and deflected beams. While this significantly degrades the resolution of our slit flat fields, the flat only varies slowly as a function of color and its change across the degraded resolution is negligible. The resolution of our science images is set by the fiber size, not the slit width and therefore widening the slit has no impact on our science images.
 
\begin{figure*}
    \centering
    \captionsetup[subfigure]{position=top, labelfont=bf,textfont=normalfont,singlelinecheck=off,justification=raggedright} 
    \subfloat[\label{fig: slitflat order 2}]{\includegraphics[width=0.45\linewidth, trim=0cm 0cm 0cm 2cm, clip]{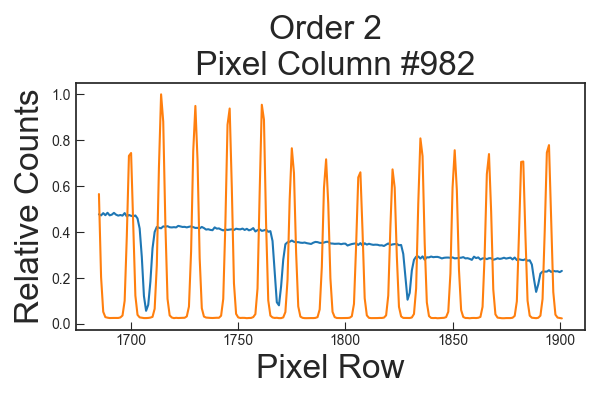}}
    \subfloat[\label{fig: slitflat order 19}]{\includegraphics[width=0.45\linewidth, trim=0cm 0cm 0cm 2cm, clip]{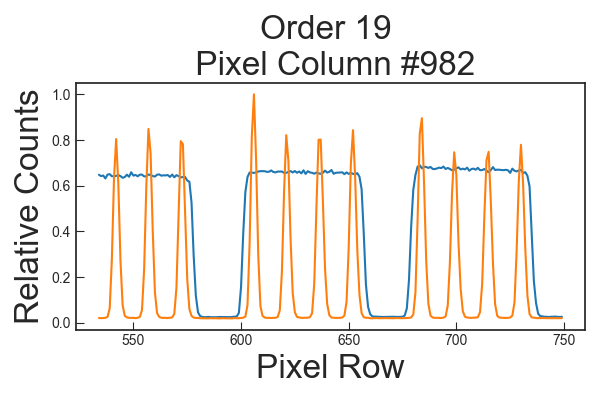}}
    \caption{The blue lines show a normalized cross section of the slit flat calibration image. In orange, we overlay the same cross section of the normalized traces in the science frame taken the same day, showing that the slit flats give us adequate signal to calibrate the pixel to pixel variations under the science fibers. Figure \ref{fig: slitflat order 2} shows the most crowded orders at the blue extreme, and Figure \ref{fig: slitflat order 19} shows the least crowded orders at the red extreme, showing that we have adequate signal and sufficient separation at both extremes.
    \\(A color version of this figure is available in the online journal.)}
\end{figure*}

Figures \ref{fig: slitflat order 2} and \ref{fig: slitflat order 19} show the cross section of the bluest and reddest orders, respectively, of the slit flat (without the iodine cell) overlayed on the same cross section of a daytime sky spectrum (with the iodine cell) to show that the slit flats give us adequate signal to calibrate the pixel to pixel variations under the science fibers, despite the deflection of the science fibers due to the iodine cell.

\section{Spectrograph environmental performance}
\label{sec:stability}

\begin{figure*}
\centering
\captionsetup[subfigure]{position=top, labelfont=bf,textfont=normalfont,singlelinecheck=off,justification=raggedright} 
\subfloat[\label{fullP}]{\includegraphics[width=8.5cm, trim=0cm 0cm 0cm 1cm, clip]{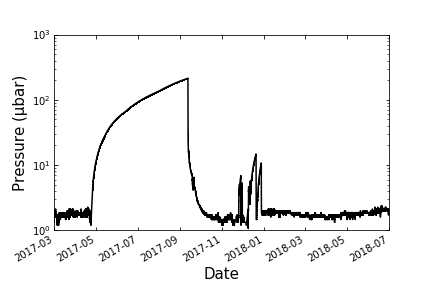}}
\subfloat[\label{monthP}]{\includegraphics[width=8.5cm, trim=0cm 0cm 0cm 1cm, clip]{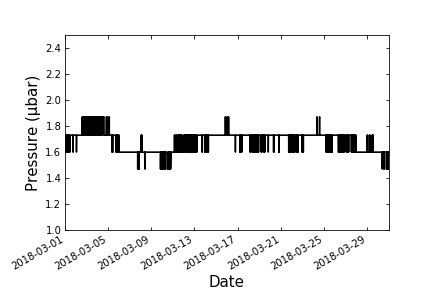}}
\vspace{0mm}
\caption{Spectrograph pressure inside the KiwiSpec spectrograph. The plotted measures were taken between 2017 March and 2018 July. A power outage at the MINERVA facility resulted in the pressure spike seen during 2017 mid-April. Alongside is plotted the month of 2018 March to show the shorter term stability after this was resolved. Note the pressure shown here is quantized because we are approaching the limit of our Granville Phillips 275 Convectron Gauge.}
\label{fig:pressure}
\end{figure*}

Here we show the pressure stability of the spectrograph (Figure \ref{fullP}) from 2017 March through 2018 July at 3~s intervals. It should be noted however that during 2017 mid-April, a power outage occurred at the MINERVA facility. Typically, the spectrograph is continuously pumped, but the outage caused the valve from the pump to the spectrograph to close, which went unnoticed for an extended period of time (mostly during the monsoon when we were not operational). We have since implemented a watchdog that sends an email notification if the pressure rises above 10~$\mu$bar (see \S \ref{sect: software}). This power outage resulted in a swift increase in pressure as it leaked up toward atmosphere. By early October, the pressure once again became sufficiently stable for the collection of good quality data, and remained so through 2018 July aside from some minor fluctuations due to maintenance. Figure~\ref{monthP} shows this stability over the month of 2018 March with a RMS of 0.065~$\mu$bar (dramatically exceeding our requirement of 12~$\mu$bar).

\begin{figure*}
\centering
\captionsetup[subfigure]{position=top, labelfont=bf,textfont=normalfont,singlelinecheck=off,justification=raggedright} 
\subfloat[\label{fullT}]{\includegraphics[width=8.5cm, trim=0cm 0cm 0cm 1cm, clip]{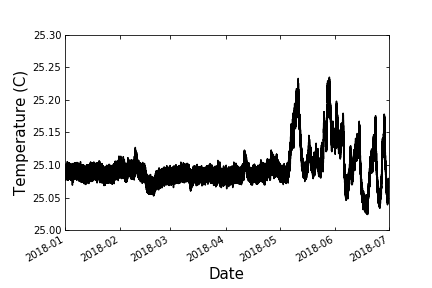}}
\subfloat[\label{monthT}]{\includegraphics[width=8.5cm, trim=0cm 0cm 0cm 1cm, clip]{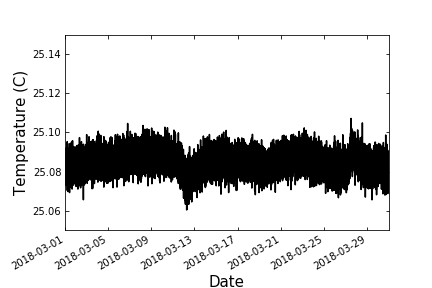}}

\vspace{0mm}
\caption{Temperature reading taken at the echelle side location on MINERVA. Measurements were taken from 2018 January through July, with small fluctuations occurring from May onward. We also plot 2018 March to more directly show the stability during a typical month of operation.}
\label{temp}
\end{figure*}

Temperature readings meanwhile, were recorded for MINERVA over the period of 2018 January through July (Figure~\ref{fullT}). These measurements were taken at the side of the cell holding the Echelle grating, and so it is most relevant for RV stability. Throughout that time period we manage to remain fairly stable from January to May, one such example being Figure~\ref{monthT} which depicts the stability over the month of March, with an RMS of 0.0052 K (two times better than our requirement of 0.01 K). Slightly larger fluctuations were seen to occur from May onward, where we removed and reinstalled the outer thermal enclosure for maintenance.

Moving forward, we intend to investigate the causes of some of the more minor fluctuations present in the environmental data. These could be a result of events such as the backlight being turned on and off, moving of the iodine stage in and out (which holds the cell which must be heated to 55~$^\circ$C), fluctuations of the room HVAC, or other events related to the operation of MINERVA itself. 

The scatter in the empirically determined wavelength solution of a single chunk for all our targets on sky (0.003~\AA, or about 0.1~pixels, for the zero-point and 0.07\% for the dispersion) dominates any trends seen on nightly or monthly timescales.

\section{Telescope Control Software}
\label{sect: software}

\subsection{Architecture}

In \citet{Swift.2015}, we described robotic control software based on the Robo-AO control software written in C \citep{Riddle.2012}. In the following years, we determined that the growing popularity of Python, the many easily importable libraries, and vendor-provided APIs made it an attractive alternative to write and maintain the code while simultaneously allowing more complex features and capabilities. Our entire operational code base, written in Python, is hosted on Github\footnote{\url{https://github.com/MinervaCollaboration}}.

A computer called ``main'' runs a 64 bit Ubuntu operating system and is responsible for most of the high-level operations. On startup, it begins three continuous functions. First, it operates an NTP server to which all other MINERVA clocks sync. It syncs itself to one of several stratum 2 time servers in the Western United States. Second, it runs a watchdog routine to monitor the temperature of the spectrograph in many locations and alert us via email if any are out of their operating range. Third, it runs a ``domeControl'' daemon that monitors the weather from several local weather stations: one at the MEarth building about 300~m away \citep{Irwin.2009}, one at the HAT building about 230~m away \citep{Bakos.2002}, one at the FLWO 1.2 m robotic telescope (home of KeplerCam) about 60~m away \citep{Szentgyorgyi.2005}, and one we installed at the MINERVA building. It automatically evaluates if it is safe to open the domes based on cloud coverage, rain, humidity, wind speed, and Sun altitude, allowing overrides to open during cloudy weather or during the day for engineering. If it has been below freezing and wet (i.e., a potential for ice or snow), it sends us an email notifying us that manual approval is required to open the domes. Snow or ice on the roof can fall on the telescopes or overload the motors that open the shutters, which can prevent further robotic or remote control. Snow at the base of the enclosure can prevent it from fully opening. It must either melt or be cleared by the local site staff before we can safely open.

All automated safety checks must pass for 30 consecutive minutes and it must be requested to open before it will actually open. Once the enclosures are open, the criteria for closing are somewhat looser. These two requirements prevent rapid cycling of the enclosure during marginal conditions. 

The domeControl daemon runs through its safety checks, sends a ``heartbeat,'' and updates a status file every 15~s. The heartbeat is a firmware-level safety feature that protects us against a variety of potential failures. If a minute has elapsed and the enclosure has not received a heartbeat, it will automatically close, independent of any other activities. Should a failure of some kind prevent the enclosure from closing, emergency text messages are sent to several people to investigate immediately.

Finally, the ``main'' machine orchestrates the observations, which start each night at 4 pm local time via cron job, which we will describe in detail later in the next subsection.

Each of the four telescopes is controlled by its own computer running a 64 bit Windows 7 Professional operating system. Windows is required to run MaximDL for camera control and the PlaneWave Interface (PWI) software for telescope control. MaximDL controls our Andor and Apogee imagers and filter wheels for photometry \citep[see][]{Swift.2015}, as well as our ZWO imagers (see \S \ref{sect:zwo}). We wrote our own server that runs locally on each Windows machine and can relay commands from our main control computer on the network to MaximDL. All images are saved to their own control computer on a drive that is cross mounted on the main computer. This allows us to run more complex image analysis like Source Extractor \citep{Bertin.1996} and astrometry.net \citep{Lang.2010} to perform automated world coordinate solutions for acquisition and guiding or automatic exposure time adjustment during sky flats.

PWI hosts its own server that can be controlled by any computer on the network through simple XML commands. PlaneWave provided several example functions in Python, which we integrated into our software. Each windows computer has ``scheduled tasks'' (the Windows equivalent to a cron job) that reboot the computers daily and start the servers.

The spectrograph is controlled by two additional Windows 7 computers provided with the KiwiSpec spectrograph from KiwiStar Optics (a business unit of Callaghan Innovations). One computer is dedicated to the thermal control servo that maintains the spectrograph temperature and runs independently of all others. 

The other computer operates much like the telescope control computers, and runs our server to relay commands from the main computer to the hardware connected directly to the spectrograph control computer. While KiwiStar Optics provided software to control the spectrograph manually, there was no API to interact with it robotically. We wrote our own spectrograph control software to enable robotic operations. This computer is responsible for operating the iodine stage, iodine heater, the spectrograph detector, the backlight, the flat field lamp, the exposure meter, and the vacuum pump and valves. The server also doubles as a watchdog that emails us if the vacuum pressure goes out of its operating range.

\subsection{Operations}

The observations begin at 4 pm local time. Our software computes the time it takes for a standard suite of biases, darks and flats for the spectrograph necessary for calibrating our RV observations. If photometric observations are desired, we upload a schedule file that contains the observations and corresponding calibrations. At 4 pm, the software computes how long the requested photometric calibrations will take, then begins the calibrations so they will finish 10 minutes before sunset.

Under normal spectroscopic observations, a dispatch scheduler reads active targets tabulated in a Google spreadsheet, and computes a score for each target based on the current time, the target's visibility, when it was last observed, how many times it has been observed that night, and how many times we would like it to be observed each night. In addition, it computes a weight for a single B-star observation that grows throughout the night until it is observed to ensure we obtain one B-star observation per night for calibrations. Further details about the MINERVA's scheduler can be found in Johnson et al. (2019 in preparation).

When photometric observations are requested for any subset of telescopes, we schedule our RV observations around the allocated times for the allocated telescopes. When only a subset of telescopes are scheduled for photometric observations, the others continue obtaining RV observations. Each telescope within the subset is capable of observing a distinct photometric target while the other telescopes obtain RVs on a single target.

During a typical spectroscopic observation, with all telescopes that are in RV mode, we slew to the target, turn on the backlight inside the spectrograph to illuminate the fibers, and expose the FAU camera. This provides us with the precise pixel location of the fiber on the acquisition camera. Next, we do a fine acquisition to put the star onto the fiber. Because our targets are so bright, it is a safe assumption to move the brightest star in the field to the position of the fiber. Then we perform an autofocus and begin guiding to keep the star onto the fiber. While we use an Alt/Az telescope, we do not correct for field rotation, opting to keep the target star on the fiber and letting all other stars rotate about it to minimize the change in the pupil illumination during the exposure.

During a typical photometric observation, we can either cycle through a list of filters throughout some observing window (e.g., a predicted transit window), observe continuously in one filter, or take some number of exposures in each of several filters. While we have an off-axis guider for the imager, MaximDL does not allow us to control three cameras simultaneously, nor does it provide an API to switch between them robotically. However, the tracking performance of the CDK700s is superb and the direct drive motors have no periodic error, allowing us to take 5 minutes exposures unguided without any measurable trailing. Therefore, instead of the off-axis guider, we use the previous science image to correct for any long-term drift in tracking. This also has the advantage of not being subject to flexure or differential field rotation between the off-axis guider and the science camera, allowing us to easily maintain sub-pixel guiding accuracy throughout an hours-long transit---a capability that is critical to obtaining precise differential photometry.

We observe either RV targets or photometric targets throughout the night as desired, all the while monitoring the status of the dome and pausing if it closes. At the end of the night, we perform another set of calibrations, and close the dome. The data are backed up to our local RAID6 NAS (which can suffer two simultaneous drive failures without data loss), and our spectroscopic reduction pipeline is initiated at 10 am local time.

\section{Spectroscopic Data Reduction}
\label{sec:1d}

The first step in our spectroscopic data reduction is to calibrate the science exposures of our spectrograph's CCD.  We collect and median stack eleven frames each night for the bias, dark current, and slit flats. For each science exposure, we subtract the overscan from the raw exposure.  We experimented with dark current subtraction but omit this in our present pipeline because the corrections are negligible and it only serves to increase the noise.  We re-normalize each bias-corrected exposure by the stacked slit flats, similar to the procedure developed in \citet{Bernstein.2015}, although we retain the blaze function.  Finally, we interpolate between fiber bundles to estimate and subtract scattered light. 

With our calibrated science frames, we are prepared to extract the one-dimensional spectrum from the two-dimensional CCD exposure.  We wrote a custom pipeline using the optimal extraction algorithm \citep{Horne.1986, Piskunov.2002, Zechmeister.2014, Bernstein.2015}.
Optimal extraction requires that flux is a separable function of $x$ and $y$ so that $F(x,y) = F(x)F(y)$, a condition that is very nearly satisfied in our instrument.  This allows us to independently find the observed flux in each row, $x$, through
\begin{equation}
  F(y) = \bm{p}(x, y)F(x) + n(y).
  \label{eqn:oe}    
\end{equation}
Here $F(x)$ is the underlying spectrum we wish to extract.  We determine this from the observed flux in the cross-dispersion direction, $F(y)$, a model for the noise $n(y)$, and a normalized cross-dispersion profile, $\bm{p}(x,y)$.

Our pipeline presently uses a modified Gaussian for $\bm{p}(x, y)$. This gives us the form
\begin{equation}
    \bm{p}(x, y) = N(x)\, e^{\textstyle \left(-0.5\left(\frac{|y-y_c(x)|}{\sigma(x)}\right)^{p(x)}\right)}.
    \label{eqn:mod_gauss}
\end{equation}
The free exponent $p(x)$ is slightly broader than a typical Gaussian with $p\approx2.2$.  The value $N(x)$ is a numerically determined normalization coefficient and $y_c(x)$ indicates the trace centroid, determined during calibration from archival fiber flats. We model $\sigma(x)$ and $p(x)$ as slowly varying polynomials along the dispersion direction

We simultaneously extract all fibers within each column, accounting for any cross-talk.  During extraction, we include a slowly varying background term to account for any additional scattered light.  We also apply a cosmic ray rejection algorithm and mask any hits.  Although our precise wavelength solution $\lambda(x)$ is found with the Doppler pipeline, we generate an initial solution from archived thorium argon exposures we took during the installation and maintenance of the spectrograph. This allows the subsequent code to quickly lock on to the correct solution.

The spectra from individual telescopes are extracted and modeled independently all the way through to the orbital modeling. This gives us four data points per exposure and a unique insight into systematic errors having to do with the telescope and the position of the trace on the detector (cosmic rays, scattered light, and flat-fielding).

\section{Doppler Pipeline}\label{sect: doppler}

The one-dimensional spectrum is the primary input for our Doppler code\footnote{\url{https://github.com/MinervaCollaboration/minerva-pipeline}}.  The architecture and general principles of our Doppler code are inspired by the code that is comprehensively described by \citet{Wang.2016}, although the algorithm is originally introduced by \citet{Butler.1996}.  Our code implements a forward modeling procedure on this spectrum that can be summarized mathematically as 
\begin{equation}\label{eq: flux}
    F_{obs}(x) = [F_{I_2}(\lambda(x)) \times F_{\star}(\lambda^{'}(x))]*\textrm{IP}(x),
\end{equation}
where $x$ is the pixel position in the dispersion direction, $\lambda(x)$ is the wavelength solution, $\lambda^{'}(x)$ is the Doppler-shifted wavelength solution, $F_{\rm obs}$ is the one-dimensional spectrum extracted from our observations, $F_{I_2}$ is the normalized absorption spectrum of our iodine cell, $F_{\star}$ is the stellar flux, and IP$(x)$ is a model of the spectrograph's intrinsic instrumental profile (which is sometimes referred to as the spectrograph response function or the one-dimensional spectral point spread function).  After determining the product of the iodine absorption spectrum and stellar spectrum, the observed spectrum is modeled as this product convolved with the instrumental profile. 

\subsection{Iodine Absorption Spectrum}

We obtain $F_{I_2}$ from a high resolution Fourier Transform Spectrometer (FTS) scan of the gaseous iodine cell as it is illuminated by a high signal-to-noise-ratio (S/N) continuum light source.  We have two FTS scans of the MINERVA iodine cell. The first one was obtained at the Pacific Northwest National Laboratory a few years ago together with the CHIRON iodine cell~\citep{Tokovinin.2013}. The second FTS scan was done by Dr. Gillian Nave's group at NIST~\citep[e.g., see][]{Nave.2011, Crause.2018}.  Both scans were taken with the iodine cell at its operating temperature specification of 55~$^\circ$C. Unfortunately, the two scans disagree in terms of line depths and line depth ratios, and we are further investigating the origin of this discrepancy. For concreteness, the results shown here use the second FTS scan, though both produce similar results.

The FTS scan is sufficient for determining our fiducial wavelength solution $\lambda(x)$ because the resolving power of the FTS ($R \approx 300,000$) is about a factor of 4 greater than that of our KiwiSpec spectrograph ($R\approx 80,000$).  Because the molecular iodine lines span from 500 to 630~nm, the wavelength solution is determined solely within this range.

\subsection{Choice of IP$(x)$}

In our Doppler code, we choose between two functional forms for our model (IP$(x)$) of the spectrograph's instrumental profile.  One form is a time-invariant spline function that is introduced in \S\ref{sect: fixed} and the other is a time-varying summation of satellite Gaussian profiles stacked on one central Gaussian profile that is described in \S\ref{sect: sumgaussians}.  When using the former, we characterize our instrumental profile by using observations of a continuum light source with a high S/N while it illuminates the iodine gas cell in our KiwiSpec spectrograph.  For reasons discussed later, we use the scattered sunlight of the daytime sky as our light source.  For the Gaussian-like IP$(x)$, however, we characterize the instrumental profile simultaneously with each stellar spectrum during our forward modeling. As a precaution, we also observe a B-type star each night (with iodine cell in place) to allow a more precise characterization of our instrumental profile as it changes over long periods of time. 

\subsection{Reference Stellar Spectrum}

A reference stellar spectrum is needed to determine the magnitude of the Doppler-shift seen when observing the science target.  We use reference stellar spectra previously constructed by the CPS group using Keck/HIRES.  They find the references by observing the science target without contamination from the iodine gas absorption lines.  In this case, the IP$(x)$ can be deconvolved with this observed (iodine-free) spectrum to get a reference stellar spectrum.  In other words, $F_{\star}(\lambda_{\rm ref}(x))$ is solved for via $F_{\rm obs,ref}(x) =  F_{\star}(\lambda_{\rm ref}(x)) *\textrm{IP}(x)$.  This IP$(x)$ and $\lambda_{\rm ref}(x)$ here however are found using observations taken immediately before and after the iodine-free observation.  The iodine-free observation of the science target is bracketed by iodine-calibrated observations of a nearby B-type star.  The bracketed observations are particularly helpful if the instrumental profile is known to fluctuate on very short timescales, which is true in the case of Keck/HIRES.  An IP$(x)$ and $\lambda_{\rm ref}(x)$ is evaluated for each iodine-calibrated B-type star observation and subsequently averaged.  The resultant IP$(x)$ is then deconvolved with the spectrum from the iodine-free science-target observation $F_{\rm obs,ref}(x)$ to get $F_{\star}(\lambda_{\rm ref}(x))$, whose wavelength solution is assumed to be the averaged $\lambda_{\rm ref}(x)$.  The aforementioned CPS group refers to the reference stellar spectrum as the Deconvolved Stellar Spectral Template (DSST). 

Using these DSSTs may limit our ability to accurately model $F_{\rm obs}(x)$ because the two spectrographs may suffer from different systematics.  Furthermore, the observatory locations (Hawaii and Arizona in the U.S.) have significantly different water columns, dramatically changing the telluric features in our spectra. These differences may be a source of systematic error in our forward modeling procedure via the DSSTs.  We will investigate the extent of these errors in the future.  Meanwhile, we find that the DSSTs are sufficient for the first RV results of our RV survey.  We have yet to derive our own reference stellar spectra because their development requires a substantial amount of observing time for each of our targets. The DSSTs are derived from observations with a higher S/N, and Keck's large aperture allows it to obtain such high S/N observations in a shorter time compared to MINERVA, which minimizes complications due to barycentric motion. In addition, unlike our fiber size, Keck's slit width is adjustable, allowing higher resolution templates which is helpful in developing the template.

\subsection{Doppler-shifted Wavelength Solution}

The forward modeling procedure finds the best fit to $F_{\rm obs}(x)$ in Equation~\ref{eq: flux}.  Our Doppler code uses a least-squares algorithm to evaluate the best fit parameters.  One of the parameters is the Doppler shift of the stellar spectrum $F_{\star}(\lambda^{'}(x))$.  The Doppler-shifted wavelength solution can be decomposed as $\lambda^{'} = \lambda_{\rm ref}\cdot(1+z)$, where $z$ is the Doppler shift from the radial velocity of the star and the motion of the telescope with respect to the star. This relative motion of the telescope is our barycentric velocity and it is dominated primarily by the Earth's rotation ($\sim 0.5$~km~s$^{-1}$) and orbital motion around the Sun ($\sim 30$~km~s$^{-1}$).  The methods introduced in \citet{Wright.2014} are used in our Doppler pipeline to correct for the telescope's barycentric velocity and subsequently determine the radial velocity of the star -- which may or may not contain information about a planetary companion.  

\begin{figure} 
    \centering
    \includegraphics[width=\linewidth, trim=0in 0in 0in 0in, clip=True]{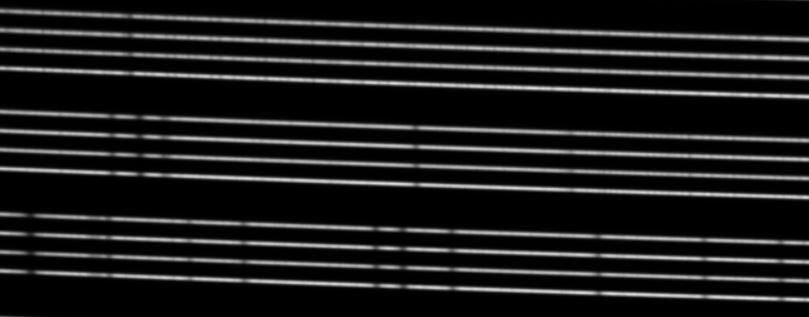} 
    \\[0.2\baselineskip] 
    \includegraphics[width=\linewidth]{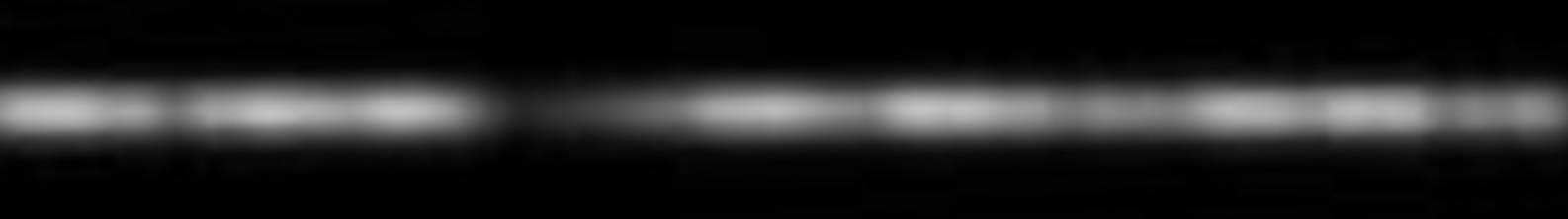}
    \caption{\label{fig: raw spectra} Top: snippet of MINERVA's two-dimensional echellogram.  Each order contains 4 traces.  Each trace belongs to each telescope.  The full width of a frame is 2048 pixels.  Bottom: a close-up of one chunk of one trace in one order is shown.  A chunk spans 128 pixels in the dispersion direction and $\sim$10 pixels in the cross-dispersion direction.  The 128 pixels translates to $\sim$2~\AA for our KiwiSpec spectrograph.  Each column of the chunk is treated as an independent crosscut of data.  The dark regions along this chunk indicate the presence of absorption features.}
    
    \vspace{5pt}
\end{figure}

\subsection{Fixed IP and IP Stability}\label{sect: fixed}

Our optical system can be divided into three general components: the telescope, the optical fibers, and the spectrograph.  As the stellar rays trace this path, the optics distort the star's image.  To determine the radial velocities from our observed spectra, we must know the manner of distortion that occurred en route to the spectrograph's CCD.  The IP$(x)$ is the shape a delta function would have when distorted by the entire optical system. The convention for determining an IP$(x)$ is to assume it follows some function comprised of Gaussian structures that extend in the dispersion direction. This section describes how we have taken a unique approach.

In developing our new IP$(x)$, we do not use a B-type star nor do we use calibration lamps.  Instead, we take spectra of the sky during the daytime.  Unlike our nightly stellar spectra, these daytime sky spectra yield a $\textrm{S/N/pixel}>150$. With our stable spectrograph, the IP$(x)$ characterized from daytime sky spectra should not change by the time we take stellar spectra at night. While the daytime sky is uniformly illuminated and may mask IP variations due to imperfect scrambling, such spectra give us a starting point to evaluate IP variations due to changes beyond the fiber.

The daytime sky spectra have proven to be a reliable source of data for determining the time scales at which our instrumental profile is stable.  To find this time scale, we create an IP$(x)$ that is time-invariant, which we refer to as the ``fixed IP." By deducing the instrumental profile in these spectra, we can determine when and why it evolves. 

We use the profile in the cross-dispersion direction to model the instrumental profile's shape in the dispersion direction. While the circular fiber makes this approximately correct, the distortions caused by the spectrograph's optical design certainly invalidate this assumption in detail. However, we can still evaluate the stability of the IP regardless, and we suspected that as long as the IP was stable and systematically wrong the same way each time, it would not impact the RV precision. Indeed, the results of long-term stability described later in this section justify this assumption.

The two-dimensional echellogram from MINERVA has four traces per order as shown in Figure~\ref{fig: raw spectra}. There is a trace for each telescope, and we divide each of the 18 orders into 15 ``chunks."  Each chunk consists of 128 columns of the trace.  The total number of chunks in a frame is the number of chunks per order times the number of telescopes times the number of orders in the frame---1080.  We define a distinct IP$(x)$ for each of these chunks because the length of each chunk acts as a characteristic length scale for which the intrinsic instrumental profile changes.  For this reason, we apply Equation~\ref{eq: flux} only over 128 pixels in the dispersion direction. Thus in practice, our forward modeling procedure is repeated for each chunk.

\begin{figure*}[t]
\centering
\captionsetup[subfigure]{position=top, labelfont=bf,textfont=normalfont,singlelinecheck=off,justification=raggedright}
\subfloat[\label{fig: 1 day IP}]{\includegraphics[width=0.329\linewidth]
{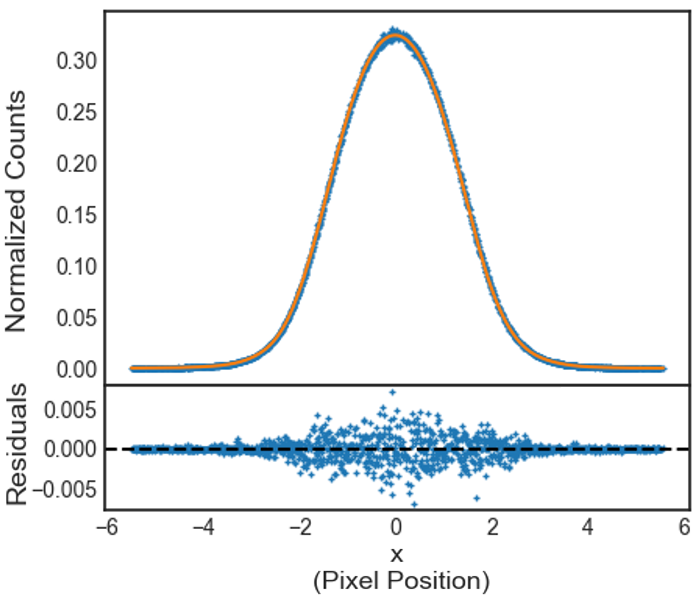}}\quad
\subfloat[\label{fig: X months IP}]{\includegraphics[width=0.31\linewidth,trim=0.8cm 0cm 0cm 0cm, clip]
{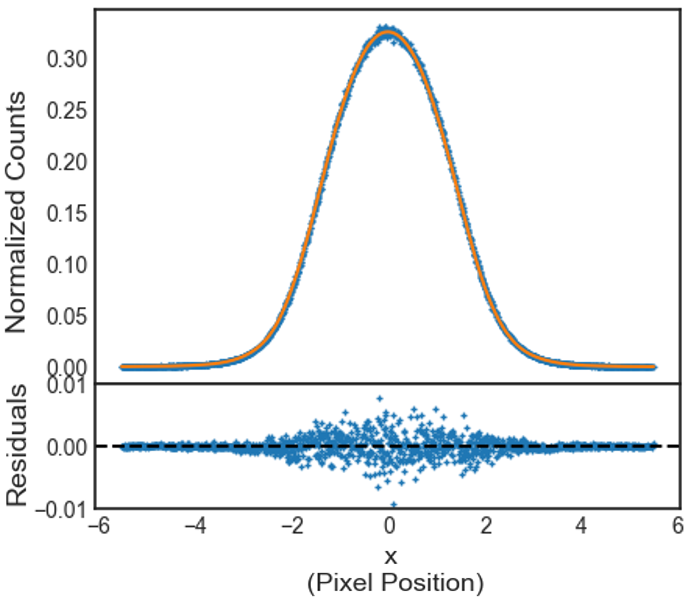}}\quad
\subfloat[\label{fig: changed IP}]{\includegraphics[width=0.31\linewidth,trim=0.8cm 0cm 0cm 0cm, clip]
{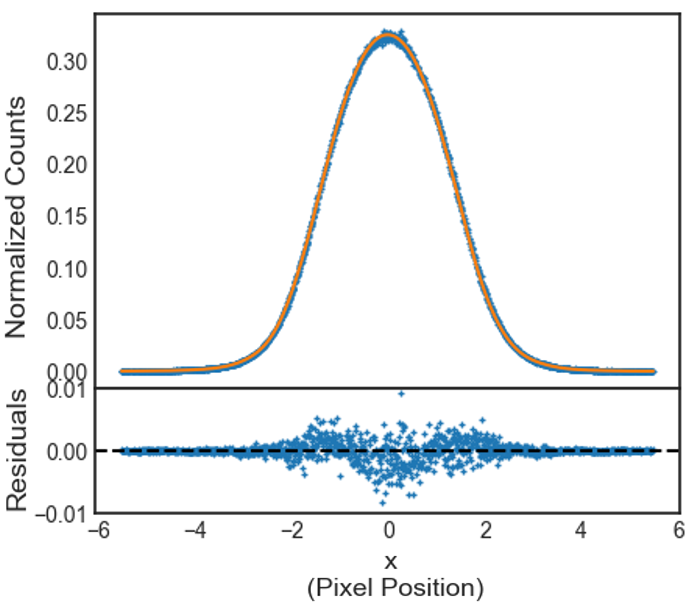}}
\caption{
Orange lines represent the fixed IP we constructed from daytime sky spectra obtained on 2018 June 19.  Blue points represent the normalized crosscuts from data obtained during one daytime sky exposure.  The bottom plots show the residuals between the normalized crosscuts and the fixed IP.  \textbf{(a)} The normalized crosscuts used to determine the fixed IP are introduced here.  \textbf{(b)} The fixed IP is used to model data obtained on 2017 October 2.  The same chunk from (a) is used here.  \textbf{(c)} The fixed IP is used to model data obtained on 2017 April 3.  The same chunk from (a) and (b) is used here.
}

\vspace{5pt}
\end{figure*}

To characterize the fixed IP, we first split a chunk (of two-dimensional daytime sky spectra) into 128 columns, or ``\nohyphens{crosscuts}."  As shown at the top of Figure~\ref{fig: raw spectra}, the traces are not perfectly horizontal.  We therefore find the centroid of each crosscut and align the crosscuts' centroids so that the chunk is essentially as horizontal as the bottom of Figure~\ref{fig: raw spectra}.  Ultimately, we want to normalize these crosscuts such that they collectively constrain the shape of our true instrumental profile.  To align and normalize them, we start by assuming each crosscut can be modeled as a Gaussian,
\begin{equation}
    G(x) = A \textrm{ }\textrm{exp}\bigg[- \bigg(\frac{x-x_0}{\sigma}\bigg)^2 \bigg] + b,
\end{equation}
where $A$ is the amplitude, $x$ is the pixel position in the cross-dispersion direction, $x_0$ is the centroid, $\sigma$ is the width, and $b$ is the background of the raw spectra.  We use a least-squares optimizer that follows the Levenberg-Marquardt algorithm to find the best fit parameters.  We then subtract $b$ from the crosscut data and integrate over this background-subtracted crosscut.  After dividing the background-subtracted crosscut by this integral we can obtain the normalized background-subtracted crosscut.  In other words, $D_{\rm norm}(x) = (D_{\rm raw}(x) - b)/N$, where $D_{\rm norm}$ is the normalized background-subtracted crosscut, $D_{\rm raw}$ is the original crosscut, and $N$ is the normalization factor calculated by the aforementioned integration.  

Once $D_{\rm norm}(x)$ is found for each of the 128 crosscuts, we fit a spline function of the third degree to all of them simultaneously.  This cubic spline has breakpoints that are each separated by 6/10 of a pixel from each other.  The spline acts as our preliminary fixed IP: IP$_{f}(x) =$~spline$(D_{\rm norm}(x))$.  To find the optimal fixed IP, we perform an iterative process of modeling the $D_{\rm raw}(x)$ and subsequently evaluating a new $D_{\rm norm}(x)$ and IP$_{f}(x)$.

Instead of a Gaussian, we use the previous spline fit model during the iterative process:
\begin{equation}\label{eq: crosscut}
    M(x) = N\times\textrm{IP}_{f}(x-\Delta x) + b,   
\end{equation}
where $M(x)$ is the model for one crosscut and $\Delta x$ is a translational shift parameter.  Now, the least-squares optimizer has only three parameters to evaluate: $\Delta x$, $N$, and $b$.  For the iterative process, we repeat the following procedure: find $D_{norm}(x)$ of each crosscut, define the IP$_{f}(x)$ for the chunk, optimize the 3 parameters of $M(x)$ for each crosscut, calculate the reduced $\chi^2$ of all the crosscuts' data and $M(x)$ models collectively, and lastly evaluate the difference between the previous iteration's reduced $\chi^2$ and the current iteration's reduced $\chi^2$.  The most important distinction between iterations is the differing fixed IPs; when the $\chi^2$ gets lower, we conclude that the current iteration's IP$_{f}(x)$ is better at modeling the spectrograph's instrumental profile than the previous iteration's IP$_{f}(x)$.  As the IP$_{f}(x)$ gets better with each iteration, the difference in reduced $\chi^2$ values lessens.  Once this difference is less than $10^{-4}$, any changes made to the IP$_{f}(x)$ in the subsequent iterations are insignificant.  The final iteration's IP$_{f}(x)$ then becomes our nominal fixed IP.

Figure~\ref{fig: 1 day IP} is an example of the final fixed IP.  The blue data points represent the $D_{\rm norm}(x)$ for all crosscuts of the final iteration.  The orange line is the final IP$_{f}(x)$.  In this case, the data come from one chunk in one daytime sky exposure taken on 2018 June 19.  The bottom plot illustrates the residuals, $D_{\rm norm}(x) - \textrm{IP}_{f}(x)$.  The residuals are greatest near the center, where the shot noise is greatest, but they show no systematic structure.  This suggests a good fit to the data.

\noindent 
\begin{table}[t]  

\begin{longtable}{ccccc}

\label{table: stars} 
\\\multicolumn{5}{c}{\scshape \tablename\ \thetable\ }\\ \multicolumn{5}{c}{ \scshape Stars of Interest }\\ \midrule\midrule  HD & R.A. (J2000) & Decl. (J2000) & $V$ & SpType \\  
\midrule  
\begin{filecontents*}{observed_stars.csv}
hd,ra,decl,v,b-v,spectype
122064,13 57 32.059,+61 29 34.30,6.52,1.01,K4V
217014,22 57 27.980,+20 46 07.80,5.46,0.70,G5V
217107,22 58 15.541,-02 23 43.39,6.16,0.76,G8IV
\end{filecontents*}
\csvreader[
		filter not equal={\hd}{217107},
		late after line=\\,%
		late after last line=\\\midrule]{observed_stars.csv}{ hd=\hd, ra=\ra, decl=\decl, v=\v, b-v=\bv, spectype=\spectype   
		}{\hd & \ra & \decl & \v & \spectype}
\end{longtable}
\vspace{5pt}
\end{table}

To test the longevity of our instrumental profile's stability for as long as possible, we construct the fixed IP with data taken at the time when we began this stability test and we used this fixed IP on spectra taken days, months, and a full year prior to the commencement of this test.  We commenced this test after the end of our first full-season observing campaign (see \S\ref{sect: status}).  We then tested the fixed IP on spectra taken as far back in time as we saw fit for this test.  We fit the same fixed IP to daytime sky spectra taken on 2017 October~2---about nine months away from the construction of the fixed IP.  The result is presented in Figure~\ref{fig: X months IP} and it has the same general pattern of noise in its residuals as Figure~\ref{fig: 1 day IP}.  This implies that the instrumental profile has not changed within that time period.  Note that for Figure~\ref{fig: X months IP} the 2018 fixed IP was used to model the 2017 spectra (via Eq.~\ref{eq: crosscut}) and subsequently normalize its crosscuts.

To extend the timeline of this test further, we tried to use daytime sky spectra taken at the very beginning of that first observing campaign (2017~September~14).  Unfortunately, our daytime sky spectra taken within those first two weeks, between September~14 and October~2 in 2017, were of poor quality and had a relatively low S/N until we resolved the issue.  Therefore, we tried to use daytime sky spectra taken before the 2017 monsoon season and thus before our first full-season observing campaign.  Fortunately for this test, we took many daytime sky spectra back in 2017 March and April.  We therefore extend the timeline of this test to 2017 April.

Figure~\ref{fig: changed IP} shows how our 2018 fixed IP is used to model data taken on 2017~April~3.  The residuals here show strong systematic structure when compared to Figures~\ref{fig: 1 day IP}~and~\ref{fig: X months IP}.  From these three examples, it is clear that our spectrograph's instrumental profile was stable from 2017~October~2 to 2018~June~19 but not from 2017~April~3 to October~2.

The spectrograph's instrumental profile evolved significantly within the window of six months between 2017 April 3 and October 2. As explained in \S\ref{sec:stability} and shown in Figure \ref{fullP}, during this time, the pressure rose dramatically for an extended period of time during the 2017 monsoon season after a brief power outage. This event permanently altered the spectrograph in such a way that the environment could not naturally return to its original instrumental profile when the pressure returned to its original operating specification. This means that the instrumental profile might have been stable for longer than nine months if the power outage and subsequent pressure instability did not occur. When characterizing the instrumental profile with a fixed IP, a new fixed IP must be used whenever an event such as this occurs. If this is not done, a situation like that of Figure~\ref{fig: changed IP} is likely to occur.

\begin{figure*}[t]
\centering
\captionsetup[subfigure]{position=top, labelfont=bf,textfont=normalfont,singlelinecheck=off,justification=raggedright}
\subfloat[\label{fig: rv standard fixed}]{\includegraphics[width=0.45\linewidth, 
	trim=0.35cm 0.45cm 0cm 0cm, 
	clip=True]{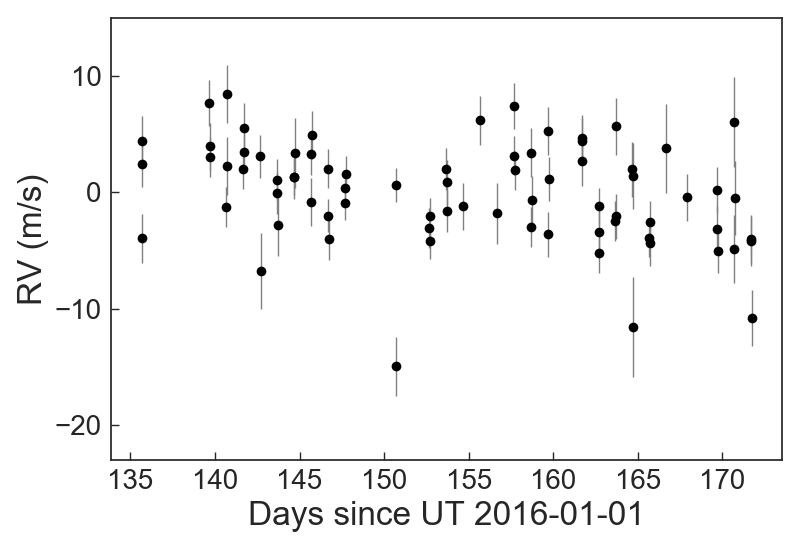}}\quad\quad
\subfloat[\label{fig: rv standard gauss}]{\includegraphics[width=0.45\linewidth, 
	trim=0.35cm 0.45cm 0cm 0cm, clip=True]{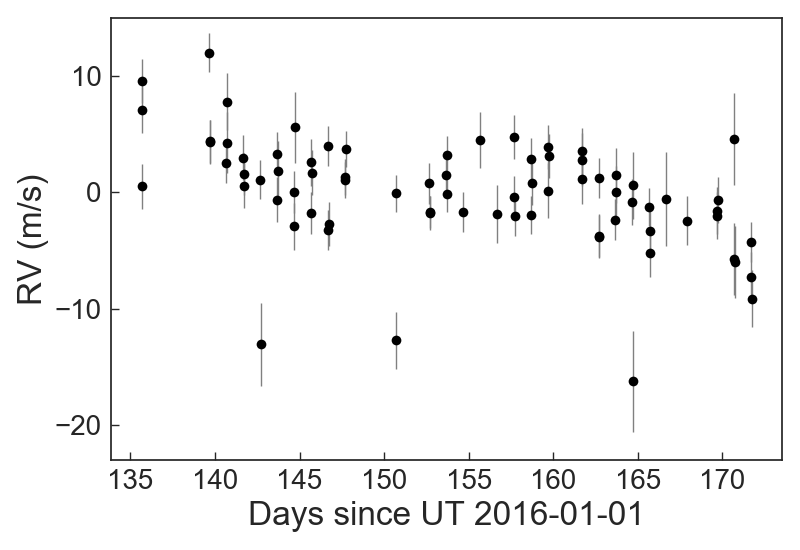}}

\caption{ \label{fig: rv standard}
MINERVA radial velocities of HD 122064.  %
\textbf{(a)} The fixed IP is used in our forward modeling procedure.  %
\textbf{(b)} Here, the sum-of-Gaussians IP is used. 
}

\vspace{5pt}
\end{figure*}

\renewcommand{\Lfctor}{0.78}
\begin{table}
\begin{longtable}{ccccc}

\label{table: RV standard star RVs}
\\\multicolumn{5}{c}{ \scshape \tablename\ \thetable\ }\\
\multicolumn{5}{c}{ \scshape HD 122064 RVs and RV Errors (m~s$^{-1}$) }

\\\midrule\midrule
Date - 2,457,500 & \multicolumn{2}{c}{IP$_f$} & \multicolumn{2}{c}{IP$_G$}
\\ (BJD$_{\rm TDB}$) & RV & Error & RV & Error 
\\ \midrule

\begin{filecontents*}{rvdata_HD122064_srt_T2.both.csv}
24.672037,-3.95,2.09,0.51,1.92
24.693414,2.47,2.03,7.07,1.99
24.714667,4.41,2.12,9.56,1.90
28.674272,7.66,1.97,12.03,1.70
28.695623,3.95,1.99,4.32,1.88
28.716867,3.05,1.75,4.39,1.88
29.677521,-1.25,1.70,2.54,1.73
29.698830,8.47,2.51,7.79,2.46
29.720215,2.27,2.51,4.25,2.62
30.675581,1.97,1.65,2.98,1.94
30.696808,3.44,1.67,1.61,1.80
30.718162,5.56,2.13,0.56,1.95
31.676155,3.09,1.84,1.10,1.69
31.697484,-6.74,3.28,-13.08,3.61
32.676528,1.07,1.78,3.34,1.85
32.697884,-0.05,1.83,-0.65,1.93
32.719243,-2.84,2.66,1.87,2.54
33.677147,1.32,1.56,0.06,1.80
33.698487,1.32,1.89,-2.92,2.01
33.719808,3.37,2.98,5.58,3.08
34.677824,-0.81,2.05,-1.76,1.82
34.699047,3.28,1.83,2.58,2.02
34.720360,4.94,2.08,1.71,1.97
35.678342,2.03,1.68,4.00,1.70
35.699559,-2.00,1.41,-3.25,1.74
35.720879,-4.01,1.79,-2.73,1.90
36.678936,-0.92,1.46,1.09,1.55
36.700295,0.38,1.47,1.35,1.56
36.721577,1.58,1.52,3.71,1.54
39.680564,0.63,1.45,-0.08,1.57
39.701867,-14.97,2.52,-12.74,2.46
41.681416,-3.05,1.75,0.78,1.77
41.702668,-4.20,1.52,-1.78,1.46
41.723934,-2.07,1.59,-1.71,1.38
42.681893,2.01,1.78,1.48,1.52
42.703186,0.86,1.95,3.22,1.60
42.724464,-1.59,1.79,-0.10,1.59
43.682388,-1.20,2.04,-1.70,1.69
44.682878,6.20,2.11,4.50,2.39
45.685934,-1.79,2.61,-1.90,2.49
46.686242,3.15,1.67,-0.38,1.78
46.707438,7.46,1.98,4.74,1.90
46.728634,1.91,1.68,-2.01,1.74
47.686578,-3.02,1.72,-1.94,1.62
47.707787,3.43,2.11,2.91,1.76
47.729072,-0.67,2.07,0.80,1.87
48.684489,-3.62,1.96,0.16,2.40
48.705826,5.27,2.10,3.88,1.89
48.727193,1.15,1.90,3.10,1.88
50.685456,4.44,1.96,3.59,1.98
50.706713,2.70,2.17,1.12,2.12
50.727985,4.65,2.03,2.83,2.29
51.685151,-5.22,1.71,-3.73,1.85
51.706517,-1.20,1.56,1.25,1.71
51.727754,-3.41,2.06,-3.80,1.82
52.685460,-2.49,1.72,-2.34,1.78
52.706773,-2.05,1.94,0.07,2.18
52.728073,5.67,2.48,1.46,2.35
53.685457,1.99,2.38,-0.80,1.98
53.706797,1.39,2.84,0.62,2.89
53.728114,-11.56,4.30,-16.26,4.36
54.685625,-3.93,1.60,-1.25,1.63
54.708278,-2.54,1.75,-3.29,1.86
54.733946,-4.37,1.93,-5.18,2.07
55.689038,3.79,3.85,-0.60,4.03
56.939036,-0.43,2.02,-2.44,2.08
58.687585,-3.15,2.00,-1.56,2.01
58.709186,0.21,1.96,-2.01,2.01
58.730659,-5.02,1.88,-0.66,1.98
59.689182,6.02,3.87,4.56,3.96
59.710383,-4.90,2.91,-5.73,3.08
59.731639,-0.49,3.19,-5.99,3.11
60.686754,-4.20,2.09,-4.25,1.73
60.709309,-4.00,2.09,-7.28,2.13
60.731261,-10.78,2.41,-9.15,2.45
\end{filecontents*}
\csvreader[head=false,
        before filter=\ifnumless{\thecsvinputline}{5}{\csvfilteraccept}{\csvfilterreject},
        late after line=\\
        ]{rvdata_HD122064_srt_T2.both.csv}{}{    \csvcoli & \csvcolii & \csvcoliii & \csvcoliv & \csvcolv
        }
        ... & ... & ... & ... & ...\\\midrule \multicolumn{5}{p{\Lfctor\linewidth}}{NOTE.---RVs for HD 122064 displayed in Figure~\ref{fig: rv standard}.  The mean error is 2.1 m/s for each data set: the data derived from the fixed IP and data from the sum-of-Gaussians IP.  \iffalse Table~\ref{table: RV standard star RVs} is published in its entirety in the electronic edition of \textit{PASP}.\fi  A portion is shown here for guidance regarding its form and content. }
        
\end{longtable}
\end{table}

\subsection{Gauss-Hermite IP}

We explored the possibility of modeling a fixed component of the IP with a time-varying component to account for changes in the IP due to perturbations like that which is seen in Figure~\ref{fig: changed IP} and explained in Figure~\ref{fig:pressure}.
We can formulate such an IP as
\begin{equation}
\begin{split}
    \text{IP}_{GH}(x) = & \sum_n A_{n} \bigg(\frac{2}{\pi \sigma^2}\bigg)^{1/4} \frac{1}{\sqrt{n!2^n}} H_n \bigg( \frac{x\sqrt{2}}{\sigma} \bigg) e^{-\big(\frac{x}{\sigma}\big)^2} \\ 
       &  + \text{ IP}_{f}(x) . 
\end{split}
\end{equation}

This function includes the sum of the products between Gaussians of amplitude $A_{n}$ and Hermite polynomials $H_{n}$.
The systematic structure seen in Figure~\ref{fig: changed IP} suggests that we may be able to model the time variable component of the IP with fewer free parameters than a purely time-variable IP, and thus preserve the signal in the spectrum to constrain the Doppler signal we care about rather than the instrumental profile. Unfortunately, the number of additional parameters required to accurately model the time-variable component was comparable to purely time-variable IP described in the following section, and therefore offers no advantage. Additionally, the GH parameterization of the IP is not as well behaved as the sum of Gaussians in our forward modeling procedure.

\subsection{Sum-of-Gaussians IP}\label{sect: sumgaussians}

Our sum-of-Gaussians IP,
\begin{equation}
    \textrm{IP}_{G}(x) = \sum_n A_{n} \textrm{exp}\bigg[-\bigg( \frac{x - x_{0,n}}{\sigma_{n}} \bigg)^2\bigg],
\end{equation}
is a time-dependent IP that is described in detail by \citet{Valenti.1995}.  Notice here that we do not include the fixed IP.  Also note that $A_{0}$ is fixed to 1 so that there is one large central Gaussian while all other Gaussian components act as small satellite Gaussians. To test this IP$_G$ against the fixed IP, we calculated RVs for an RV standard star and a planet-hosting star after forward modeling the data with each of the IPs, as described later in \S\ref{sect: RVs}.

\begin{figure*}[t]
\centering
\captionsetup[subfigure]{position=top, labelfont=bf,textfont=normalfont,singlelinecheck=off,justification=raggedright}
\subfloat[\label{fig: allan variance fixed}]{\includegraphics[width=0.45\linewidth,
	trim=0cm 0cm 0cm 0cm, clip=True]{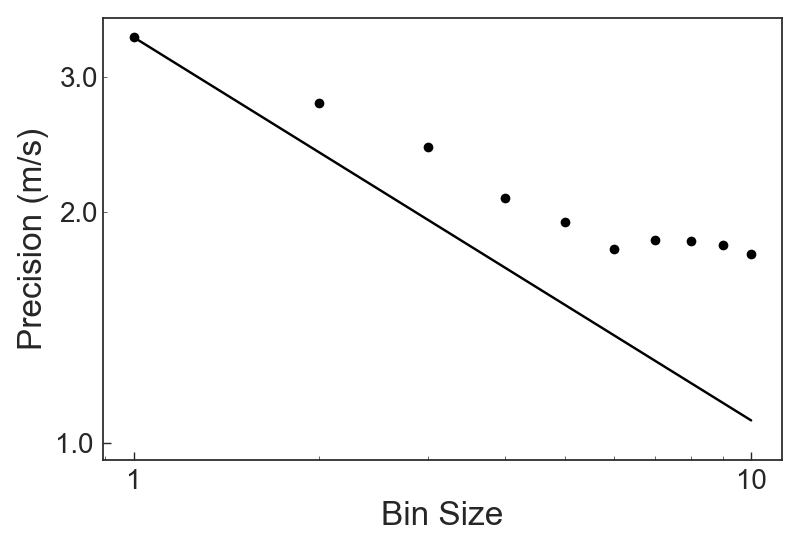}}\quad\quad
\subfloat[\label{fig: allan variance gauss}]{\includegraphics[width=0.45\linewidth,
	trim=0cm 0cm 0cm 0cm, clip=True]{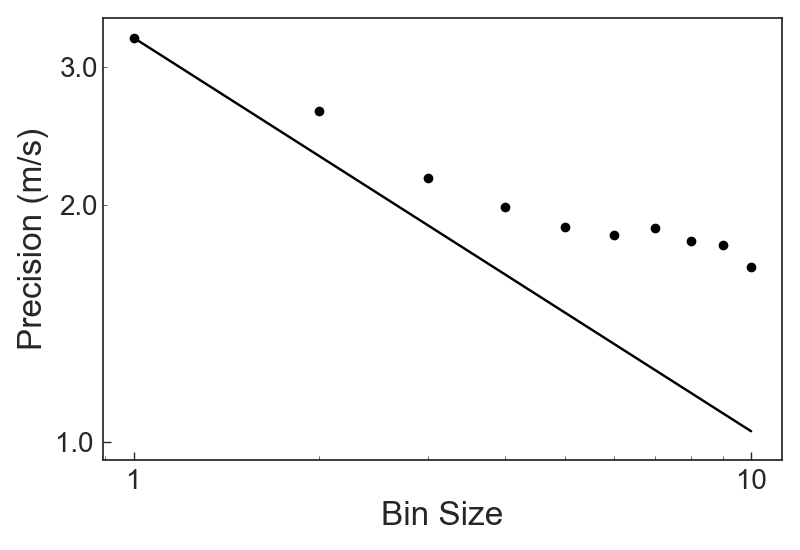}}
	
\caption{ \label{fig: allan variance}
The solid line represents the precision if the binned data set only consisted of white noise.  The points represent our precision at a given binning.  Beyond a binning of roughly six, the precision significantly deviates from the solid line, the precision barely improves, and systematic errors dominate the data.  This is the case for both the fixed IP and sum-of-Gaussians approach.  %
\textbf{(a)} Allan variance for data in Figure~\ref{fig: rv standard fixed} produced with the fixed IP.  When binning by six, we see a precision of 1.78~m~s$^{-1}$.  %
\textbf{(b)} Allan variance for data in Figure~\ref{fig: rv standard gauss} produced with the sum-of-Gaussians IP.  When binning by six, we see a precision of 1.87~m~s$^{-1}$.  
}

\vspace{5pt}
\end{figure*}

\section{RV Performance}\label{sect: RVs}

The radial velocity we measure is the reflex motion of the star induced by the gravitational pull of its planetary companion.  This motion is accounted for in Kepler's laws.  Kepler's laws suggest that the lower limit of the planetary mass can be described as
\begin{equation}  M_p \sin i \approx K\sqrt{1-e^2} \bigg( \frac{P M^2_*}{2\pi G} \bigg)^{1/3} \end{equation}
(see parameter symbols with Table~\ref{table: 51 Peg} descriptions).  To model the stellar system, we use EXOFASTv2~\citep{Eastman.2013, Eastman.2017}.  

Before the minimum mass is determined, the RV semi-amplitude must be extracted from the Doppler-shifted spectra via a forward modeling procedure.  The results of the instrumental profile work discussed in \S\ref{sect: fixed} provided fruitful information that paved the way for the successful extraction of MINERVA's first radial velocity results. 

We present RV measurements of two target stars to demonstrate MINERVA's precision. These stars are HD 122064 and HD 217014 (51 Peg), which have their coordinates, $V$ magnitude, and spectral type reported in Table~\ref{table: stars}.  HD 122064 is chromospherically inactive, has no known companions, and serves as a convenient RV standard star.  For the hot Jupiter 51 Peg b, we compare the planetary properties derived from MINERVA data with results from the literature.

\subsection{HD 122064}\label{sect: rv standard}

With one telescope, we acquired the radial velocities of HD 122064 during the months of 2016 May and June.  As a test, we used both the fixed IP and the sum-of-Gaussians IP in our forward modeling to generate two distinct RV data sets which derive from the same spectra.

For the purposes of the instrumental profile's stability test, we only used one daytime sky exposure to construct the fixed IP.  Whenever we perform our forward modeling procedure with the fixed IP, we make the fixed IP more robust by using multiple daytime sky exposures.  We perform the same procedure as described in \S\ref{sect: fixed}, except the number of crosscuts that we simultaneously fit a cubic spline function to is equal to 128 times the number of daytime sky exposures we use.  Each set of 128 crosscuts comes from the same chunk of distinct daytime sky exposures.  For the May/June data set, we use $\sim$5 daytime sky exposures to construct a fixed IP for each chunk and these exposures are somewhat evenly distributed throughout the 1.5 months timescale of the data set.  This fixed IP is used to generate the RVs in Figure~\ref{fig: rv standard fixed} while the sum-of-Gaussians IP is used to produce the RVs of Figure~\ref{fig: rv standard gauss}.

After the RVs are extracted, we compute the Allan variance to determine the level of precision MINERVA can achieve.  In Figure~\ref{fig: allan variance}, the line represents the precision if the binned data only contained white noise.  We use an error-weighted, overlapping Allan variance to determine the limit for which we can bin down the given data set before it is dominated by systematic errors \citep{Allan.1966,Malkin.2011}.  

We have seventy-five radial velocity measurements tabulated in Table~\ref{table: RV standard star RVs} and shown in Figures~\ref{fig: rv standard fixed}~and~\ref{fig: rv standard gauss}.  Figures~\ref{fig: allan variance fixed}~and~\ref{fig: allan variance gauss} suggest that a bin size of six roughly marks the limit for which the respective binned data sets begin to deviate from white noise.  At this binning, we are sensitive to variations below the 2~m~s$^{-1}$ level for our measurements of this RV standard star.  The precision achieved is 1.8~m~s$^{-1}$ for the fixed IP approach and 1.9~m~s$^{-1}$ for the sum-of-Gaussians approach.  This could potentially change depending on the standard star or the amount of data we have for a given star.  To confirm this, we plan on performing the same test for observations of other RV standard stars.  Regardless however, these RVs evaluated through our Doppler pipeline suggest that our fixed IP is doing just as well as our sum-of-Gaussians IP.

\subsection{HD 217014}\label{sect: exoplanet}

\begin{figure*}[t]
\centering
\captionsetup[subfigure]{position=top, labelfont=bf,textfont=normalfont,singlelinecheck=off,justification=raggedright}
\subfloat[\label{fig: chrono}]{\includegraphics[width=0.45\linewidth]{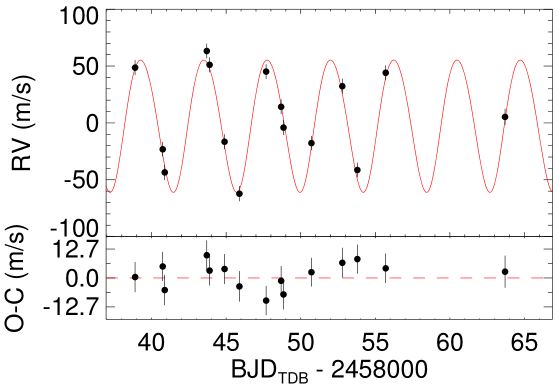}}\quad\quad
\subfloat[\label{fig: phase}]{\includegraphics[width=0.45\linewidth]{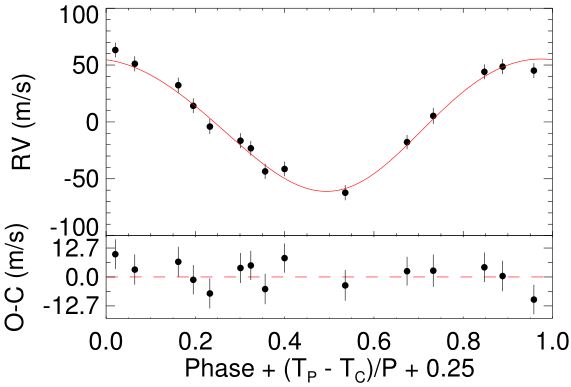}}

\label{fig: exoplanet}
\caption{
Radial velocities of 51 Peg obtained with MINERVA.  The residuals are plotted below.  The error bars listed in Table~\ref{table: 51 Peg RVs} are inflated here via the fitted jitter.  The solid line is the best-fit orbital solution derived from EXOFASTv2.  The data span across 2017 October.  %
\textbf{(a)} Radial velocity time series for 51 Peg.  Note that BJD$_{\rm TDB}$ means Barycentric Julian Date in Barycentric Dynamical Time \citep[for elaboration see][]{Eastman.2010}.  %
\textbf{(b)} The same radial velocities are phase-folded to the planet's orbital period.
\\(A color version of this figure is available in the online journal.)
}

\end{figure*}

MINERVA observations of 51 Peg were taken with one telescope in 2017 October. Again, we use the fixed IP and sum-of-Gaussians IP to extract the radial velocities.  It is wise to see if our radial velocities can confirm the existence and characteristics of exoplanet systems.  51 Peg b is the first of such exoplanets to be tested.  We use EXOFASTv2 to constrain the properties of this exoplanet system.  Our stellar parameters are informed by the broad band photometry summarized in Table~\ref{table: mags}.  The RVs and resultant 51 Peg b properties derived with both IPs are so similar that we only show the results produced with the fixed IP.  The unbinned RVs and the EXOFASTv2-generated orbital solution are illustrated in Figure~\ref{fig: chrono}, tabulated in Table~\ref{table: 51 Peg RVs}, and summarized in Table~\ref{table: 51 Peg}.  Figure~\ref{fig: phase} shows the same but with the time series folded to the phase of the planet's orbital period.

The median and 68\% confidence intervals determined using EXOFASTv2 with the MINERVA data for all parameters of the 51 Peg system are listed in Table~\ref{table: 51 Peg}.  We only employ constraints on three of the stellar parameters.  We impose a prior on the stellar metallicity ([Fe/H]) of 0.20 $\pm$ 0.07 dex from spectroscopy described in \cite{Fuhrmann.1997}.  We set the $V$-band extinction's ($A_V$) upper limit to 0.11811 magnitudes, using the dust maps from \cite{Schlafly.2011}.  Lastly, we impose a Gaussian prior on the parallax ($\pi$) of 64.65~$\pm$~0.12~mas from \cite{Gaia.2016,Gaia.2018}.  These priors, coupled with a MIST stellar evolution model~\citep{Choi.2016,Dotter.2016} and an SED model, constrain the properties of the host star.

\begin{table}
\begin{longtable}{lccc}
\label{table: mags} 
\\\multicolumn{4}{c}{\scshape \tablename\ \thetable\ }
\\\multicolumn{4}{c}{\scshape 51 Peg Magnitudes} \\\midrule\midrule 
Band & Mag. & Used & Catalog's \\ &  & Mag. Error & Mag. Error
\csvreader[head=false,
            late after line=, 
            late after last line=\\\midrule
            ]{51peg.sed_csvclear}{}{
            \ifthenelse{ \equal{\csvcoli}{BT} }{\\\midrule\multicolumn{4}{l}{\textit{Tycho-2 Catalog} \hfill\citep{Hoeg.2000}}}{}
            \ifthenelse{ \equal{\csvcoli}{J2M} }{\\\multicolumn{4}{l}{\textit{2MASS Catalog} \hfill\citep{Cutri.2003}}}{}
            \ifthenelse{ \equal{\csvcoli}{WISE1} }{\\\multicolumn{4}{l}{\textit{WISE Catalog} \hfill\citep{Wright.2010}}}{}
            \\\csvcoli & \csvcolii & \csvcoliii & \csvcoliv\ifthenelse{ \equal{\csvcoli}{VT} }{\\}{}\ifthenelse{ \equal{\csvcoli}{K2M} }{\\}{}}
\end{longtable}
\end{table}

We compare our results with the values in \cite{Butler.2006} (hereafter referred to as Bu06).  They cite the SPOCS catalog \citep{Valenti.2005} as the source for most of their stellar parameters.  The Bu06 distance $d$ to the star and its uncertainty are from the \textit{Hipparcos}\footnote{Vizier Online Data Catalog, I/239 (ESA 1997)} catalog.  \cite{Valenti.2005} suggest that the typical uncertainties for their stellar parameters amongst their catalog of stars are 0.06 dex for log $g$, 44 K for $T_{\rm eff}$, and 0.03 dex for [Fe/H].  Bu06 assumes a 10\% uncertainty for the stellar mass $M_*$.  Bu06 does not report a value for stellar radius $R_*$.  We therefore calculate this using their stated log $g$ and $M_*$ values.  The uncertainty in the stellar radius is found using propagation of error between those two parameters.  

The reference planetary and telescope parameters shown in Table \ref{table: 51 Peg} were derived solely from Bu06.  Their observations were taken at Lick Observatory using the Hamilton spectrograph~\citep{Vogt.1987}, the 3.9~m Anglo-Australian Telescope using UCLES~\citep{Diego.1990}, and the Keck Observatory using HIRES~\citep{Vogt.1994}.  Their quoted RV jitter $\sigma_J$, however, does not come from their observations.  Their jitter comes from the model developed by \cite{Wright.2005}, which was informed by a sample of 531 stars observed at Keck that had known activity levels, colors, and parallaxes.  In general, the jitter depends on the spectral type of the star and the instrument observing it.  The model by \cite{Wright.2005} uses a stellar activity indicator, $B-V$ color, and difference in magnitude above the main sequence to approximate the stellar jitter. 

\renewcommand{\Lfctor}{0.75}
\begin{table}[t]
\begin{longtable}{cccc}
\label{table: 51 Peg RVs}

\\\multicolumn{4}{c}{ \scshape \tablename\ \thetable\ }
\\\multicolumn{4}{c}{ \scshape RV Results for 51 Peg}
\\\midrule\midrule
Date - 2,458,000 & RV & Error & Residuals \\ 
(BJD$_{\rm TDB}$) & (m~s$^{-1}$) &  (m~s$^{-1}$) & (m~s$^{-1}$) 
\\ \midrule
\begin{filecontents*}{rvHD217014_Referee_orig+new.csv}
38.897636,42.28,3.11,0.41,1.45
40.748225,-29.54,2.60,5.05,3.57
40.883873,-49.87,3.19,-5.30,-6.98
43.703150,56.90,2.81,9.99,3.88
43.888337,44.76,3.13,3.24,-3.30
44.892530,-22.88,2.99,3.94,3.37
45.888237,-68.62,3.27,-3.70,-7.01
47.681088,38.81,2.72,-9.92,-13.90
48.686860,7.81,2.70,-1.24,-3.40
48.844736,-10.45,3.04,-7.23,-8.07
50.718624,-24.11,2.30,2.54,3.96
52.787278,25.94,3.05,6.67,4.13
53.797358,-47.75,2.59,8.28,7.65
55.696227,37.76,2.81,4.24,4.52
63.696364,-1.04,3.99,2.76,2.71
\end{filecontents*}
\csvreader[head=false,
            late after line=\\
            ]{rvHD217014_Referee_orig+new.csv}{}{
            \csvcoli & \csvcolii & \csvcoliii & \csvcoliv }
            \midrule \multicolumn{4}{p{\Lfctor\linewidth}}{NOTE.---The mean formal error derived from the third column is 3.0~m~s$^{-1}$.  The fourth column represents the residuals between the RV data (of the second column) and the best-fit orbital solution from EXOFASTv2. The RMS of the residuals is 5.6~m~s$^{-1}$, and the systematic error floor, achieved when binning by 2, is 4.2~m~s$^{-1}$.}
\end{longtable}
\end{table}

Values not reported by Bu06 are marked with ellipses (...) in Table~\ref{table: 51 Peg}.  For the parameters that have a value and errors reported by Bu06, we state the discrepancy between our values and theirs in terms of 1$\sigma$ uncertainty.  We define this discrepancy as 
\begin{equation}\label{eq: compare}
\Delta\sigma = \frac{ N_1 - N_2 }{ \sqrt{ (\sigma_{1,L})^2 + (\sigma_{2,U})^2 } }, \textrm{when } N_1 > N_2,
\end{equation}
in which $\sigma_{1,L}$ is the lower error bar for $N_1$ and $\sigma_{2,U}$ is the upper error bar for $N_2$.  Ideally, the discrepancy should be less than 1$\sigma$.  Seeing as the discrepancies evaluated for the tabulated parameters---most importantly the planetary parameters---are $\lesssim 1.0\sigma$, we find good agreement with results quoted in literature.

\begin{figure}
    \centering
    \includegraphics[width=\linewidth]{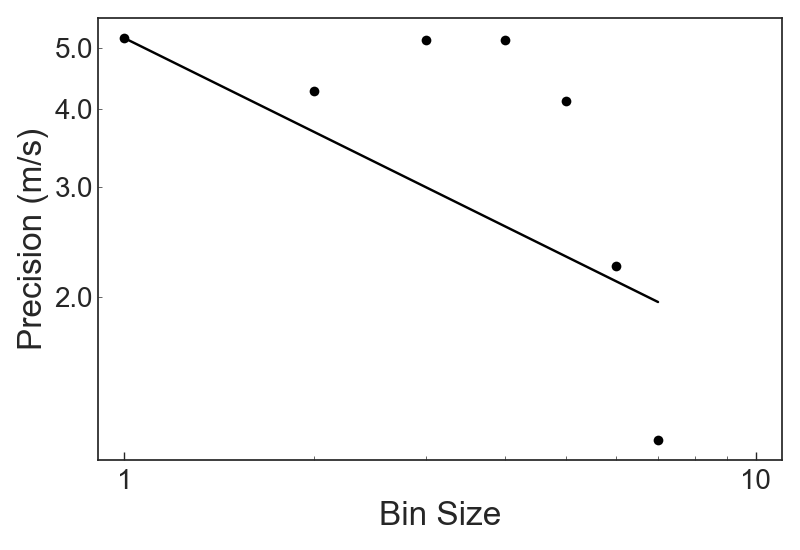}
    \caption{ Allan variance from the 51 Peg model residuals listed in Table \ref{table: 51 Peg RVs}.  The precision for the RV standard star (HD 122064) is notably better than the precision exhibited here for 51 Peg.  The additional RV scatter is most likely due to the greater stellar jitter and substantially smaller sample size of this data set. }
    \label{fig:51pegallanvariance}
\end{figure}

We also analyze the residuals of the fit to 51 Peg b to determine our precision. An Allan variance plot of the residuals is shown in Figure \ref{fig:51pegallanvariance}, demonstrating our per point scatter (5.6~m~s$^{-1}$) and systematic floor (4.2~m~s$^{-1}$) when two observations are binned is considerably worse than achieved for HD 122064, most likely due to stellar jitter of the more active host star. We note that fitting for the period and eccentricity may absorb some of the excess scatter and make our precision appear better than it is. To address this, we also ran a fit fixing the period to the Bu06 value (4.230785 days) and fixing the eccentricity to zero, but found an insignificant increase in the per-point residual RMS (5.6 to 6.0~m~s$^{-1}$) and an insignificant decrease in the binned-by-2 residual RMS (4.3 to 4.2~m~s$^{-1}$). Therefore, we conclude that fitting the period and eccentricity does not have a major impact on our inferred precision.

\section{Future Work}
\label{sec:future}
While we are about a factor of 2 of our original goal and already operating with an unmatched combination of cadence and precision that enables us to detect new planets and provide valuable insight into stellar jitter, there are several areas for improvement that may help us achieve our original goal of 80~cm~s$^{-1}$.

The relatively long exposures we typically take increases the uncertainty in the flux-weighted midpoint time and therefore introduces additional error in the corresponding barycentric correction \citep{Wright.2014}. Either shortening our exposures or improving the determination of the flux-weighted midpoint may improve our ultimate RV precision.

The stability of our spectrograph demonstrated here allows us several avenues to improve our Doppler pipeline. The templates we use, derived from Keck/HIRES observations, is likely to contain systematic differences due to the instrument and atmospheric conditions (e.g., increased water column) from our instrument, and may be dominating our RV error. We will investigate generating our own templates, but we will also investigate fitting for the template from dozens of spectra, known as the ``grand solution'' \citep{Gao.2016,Czekala.2017}. Further, modeling many spectra at once with the same instrumental profile and/or fitting for the iodine cell removes many sources of potential systematic error and unnecessary nuisance parameters that may be covariant with the radial velocity.

While we have octagonal fibers coupled to a circular fiber for optimal near-field scrambling, we have opted not to introduce a double scrambler to improve our far field scrambling, which typically reduces the throughput by 10-20\% \citep{Halverson.2015}. We may revisit this trade-off in the future, as well as explore the possibility of introducing an agitator to improve modal noise. We are also actively exploring the improvement achievable by using spectro-perfectionism to improve the extraction of our 1D spectra using a 2D instrumental profile \citep{Bolton.2010}.

Finally, having four simultaneous spectra gives MINERVA a unique insight into telescope and detector level systematics which we have yet to fully capitalize on. In particular, we will explore the possibility of adapting the ``vanking'' stage of the Doppler pipeline, which is a sophisticated outlier rejection algorithm, to incorporate the knowledge that the four simultaneous spectra should produce identical radial velocities.

\section{Conclusion}\label{sect: conclusion}

Since the commissioning of MINERVA, we have substantially modified our telescope control software and our Doppler pipeline.  The MINERVA mission's secondary goal is accomplished much more efficiently with the control software changes.  This work marks our first achievement toward our primary science goal of obtaining precise RVs.  We have confirmed which of the IPs we had at our disposal would yield reliable results from our pipeline.  These are the aforementioned time-invariant cubic spline function (the fixed IP) and the sum-of-Gaussians function.  While testing the fixed IP, we have also confirmed that our spectrograph's intrinsic instrumental profile remains stable for months.  When there is significant fluctuation in the intrinsic instrumental profile, it is likely due to disturbances to the instrument, as opposed to any natural and gradual perturbations within the instrument.  
The agreement between both IPs implies that using an instrumental profile from the cross dispersion direction, and therefore has systematic errors, is sufficient if the instrumental profile is stable. Consequently, we precisely characterized our spectrograph's instrumental profile from the cross-dispersion direction of the echellogram. 

\acknowledgements

MINERVA is a collaboration among the Harvard-Smithsonian Center for Astrophysics, The Pennsylvania State University, the University of Montana, and the University of Southern Queensland. MINERVA is made possible by generous contributions from its collaborating institutions and Mt. Cuba Astronomical Foundation, The David \& Lucile Packard Foundation, National Aeronautics and Space Administration (EPSCOR grant NNX13AM97A), The Australian Research Council (LIEF grant LE140100050), and the National Science Foundation (grants 1516242 and 1608203). Any opinions, findings, and conclusions or recommendations expressed are those of the author and do not necessarily reflect the views of the National Science Foundation.  
Funding for MINERVA data-analysis software development is provided through a subaward under NASA award MT-13-EPSCoR-0011.  
This work was partially supported by funding from the Center for Exoplanets and Habitable Worlds, which is supported by the Pennsylvania State University, the Eberly College of Science, and the Pennsylvania Space Grant Consortium.  
We are grateful to Dr. Gillian Nave and R. Paul Butler for providing FTS measurements of our iodine gas cell. 

\renewcommand{\Lfctor}{1}
\begin{longtable*}{ccccc}  
\label{table: 51 Peg}
\\ \multicolumn{5}{c}{ \scshape \tablename\ \thetable\ }
\\ \multicolumn{5}{c}{ \scshape Properties of 51 Peg }\\\midrule\midrule Parameter & Description & EXOFASTv2 & Reference (Bu06) & $\Delta\sigma$ \\ \midrule  

\endfirsthead
 
\multicolumn{5}{c}{ \scshape \tablename\ \thetable\ --- \textit{Continued }}\\\midrule\midrule Parameter & Description & EXOFASTv2 & Reference (Bu06) & $\Delta\sigma$ \\ \midrule
\endhead

\endfoot  


\endlastfoot

\begin{filecontents*}{51peg_star.csv}
parsymbol,parinfo,exofast,litval,reference,sigdiff
$M_*$,Mass ($M_\odot$),$1.095^{+0.066}_{-0.07}$,$1.09\pm 0.109$,VF05,0.039
$R_*$,Radius ($R_\odot$),$1.139^{+0.03}_{-0.03}$,$1.03\pm 0.07$,\ldots,1.431
$L_*$,Luminosity ($L_\odot$),$1.39^{+0.056}_{-0.053}$,\ldots,\ldots,\ldots
$\rho_*$,Density (cgs),$1.05^{+0.12}_{-0.11}$,\ldots,\ldots,\ldots
$\log{g}$,Surface Gravity (cm s$^{-2}$),$4.365^{+0.036}_{-0.041}$,$4.449\pm 0.06$,\ldots,1.2
$T_{\rm eff}$,Effective Temperature (K),$5875^{+73}_{-76}$,$5787\pm 44$,\ldots,1.002
$[{\rm Fe/H}]$,Metallicity,$0.201^{+0.07}_{-0.069}$,$0.2\pm 0.03$,\ldots,0.013
$[{\rm Fe/H}]_{0}$,Initial Metallicity,$0.204^{+0.064}_{-0.065}$,\ldots,\ldots,\ldots
$Age$,Age (Gyr),$4.0^{+3.4}_{-2.5}$,\ldots,\ldots,\ldots
$EEP$,Equal Evolutionary Point,$367^{+39}_{-37}$,\ldots,\ldots,\ldots
$A_V$,V-band extinction,$0.051^{+0.045}_{-0.036}$,\ldots,\ldots,\ldots
$\sigma_{SED}$,SED photometry error scaling,$3.01^{+1.1}_{-0.7}$,\ldots,\ldots,\ldots
$d$,Distance (pc),$15.468^{+0.03}_{-0.029}$,$15.36\pm 0.18$,\ldots,0.592
$\pi$,Parallax (mas),$64.65^{+0.12}_{-0.12}$,\ldots,\ldots,\ldots
\end{filecontents*}
\csvreader[
		after head= \\  \multicolumn{1}{l}{ \scshape Stellar Parameters: } \vspace{5pt}\\, 
		after line=\vspace{5pt}, 
		late after line=\\,
		late after last line=\vspace{5pt}\\]{51peg_star.csv}{parsymbol=\psymb, parinfo=\pinf, exofast=\exof, litval=\plit, reference=\reph, sigdiff=\wdiff  
		}{\psymb \dotfill & \pinf \dotfill  & \exof  & \plit & \wdiff }

\csvreader[
        before filter=\ifnumless{\thecsvinputline}{16}{\csvfilteraccept}{\ifnumgreater{\thecsvinputline}{27}{\csvfilteraccept}{\csvfilterreject}},
		before first line= {\scshape Planetary Parameters:} \vspace{5pt}\\, 
		after line=\vspace{5pt},
		late after line=\\,
		late after last line=\vspace{5pt}\\]{ 51peg_planet.csv
		}{parsymbol=\psymb, parinfo=\pinf, exofast=\exof, litval=\plit, reference=\reph, sigdiff=\wdiff  
		}
		{\psymb \dotfill & \pinf \dotfill  & \exof  & \plit\ifthenelse{ \equal{\psymb}{$e$} }{$^a$}{}\ifthenelse{ \equal{\psymb}{$\omega_*$} }{$^a$}{} & \wdiff }

\begin{filecontents*}{51peg_telescope.csv}
parsymbol,parinfo,exofast,litval,reference,sigdiff
$\gamma_{\rm rel}$,Relative RV Offset (m/s),$-5.0^{+2.5}_{-2.5}$,\ldots,\ldots,\ldots
$\sigma_J$,RV Jitter (m/s),$7.7^{+2.8}_{-1.9}$,3.7,\ldots,\ldots
$\sigma_J^2$,RV Jitter Variance,$59^{+51}_{-26}$,\ldots,\ldots,\ldots
$N_{\rm obs}$,Number of observations,15,256,\ldots,\ldots
\end{filecontents*}
\csvreader[
		before first line= {\scshape Telescope Parameters:} \vspace{5pt}\\ , 
		after line=\vspace{5pt}, 
		late after line=\\,
		late after last line=\\\midrule \multicolumn{5}{p{\Lfctor\linewidth}}{ NOTE. }\\
		\multicolumn{5}{p{\Lfctor\linewidth}}{$^a$The uncertainties reported by Bu06 for $e$ and $\omega_*$ are non-Gaussian because the uncertainty of $e$ is comparable to $e$, i.e. $\sigma_e \gtrsim e/2$.  }\\ 
		\multicolumn{5}{p{\Lfctor\linewidth}}{$^b$Our RV jitter is informed by the aforementioned radial velocities while the Bu06 value is informed merely by other stellar parameters. }  
		]{51peg_telescope.csv}{parsymbol=\psymb, parinfo=\pinf, exofast=\exof, litval=\plit, reference=\reph, sigdiff=\wdiff  
		}
		{\psymb \dotfill & \pinf \dotfill  & \exof\ifthenelse{ \equal{\psymb}{$\sigma_J$} }{$^b$}{}  & \plit\ifthenelse{ \equal{\psymb}{$\sigma_J$} }{$^b$}{} & \wdiff } 

\end{longtable*}


\end{document}